\documentclass[a4paper,11pt]{article}
\pdfoutput=1 

\usepackage{jcappub} 

\usepackage[T1]{fontenc} 
\usepackage{amssymb}
\usepackage{bm}
\usepackage{amsfonts}
\usepackage{latexsym}
\usepackage[latin1]{inputenc}
\usepackage{graphicx}
\usepackage{amsmath}
\usepackage{mathpazo}
\usepackage{amsmath}
\usepackage{mathrsfs}
\usepackage{amsthm}
\usepackage{graphicx}
\usepackage{hyperref}
\usepackage{enumitem}
\usepackage{textcomp}
\usepackage{float}
\usepackage{booktabs}
\usepackage{caption}
\usepackage{subcaption}
\usepackage{dcolumn}
\usepackage{ragged2e}
\usepackage{hyperref}
\usepackage{footmisc}
\hypersetup{colorlinks,citecolor=blue}
 
\usepackage[many]{tcolorbox} 
\newtcolorbox{boxB}{
    fontupper = \bf, 
    boxrule = 1.5pt,
     width = 18.2cm,
    colframe = black,
    colback = white,
    rounded corners,
    arc = 5pt   
    }

\usepackage{amsmath}
\usepackage{xcolor}
\usepackage{orcidlink}
\usepackage{epsfig}
\usepackage{commath}
\usepackage{tabularx}
\usepackage{multirow}
\usepackage{mathtools}
\title{\boldmath Neutron Star in Covariant $f(Q)$ gravity}

\author[a,b]{Muhammad Azzam Alwan\orcidlink{0000-0002-0558-2092},}
\author[a,c,d]{Tomohiro Inagaki\orcidlink{0000-0003-2777-4017},}
\author[e]{B. Mishra\orcidlink{0000-0001-5527-3565}}
\author[e]{and S.A. Narawade\orcidlink{0000-0002-8739-7412}}

\affiliation[a]{Graduate School of Advanced Science and Engineering, Hiroshima University,\\ Higashi-Hiroshima 739-8526, Japan}
\affiliation[b]{Research Center for Quantum Physics, National Research and Innovation Agency (BRIN),\\ Tangerang Selatan 15314, Indonesia}
\affiliation[c]{Information Media Center, Hiroshima University, \\Higashi-Hiroshima 739-8521, Japan}
\affiliation[d]{Core of Research for the Energetic Universe, Hiroshima University,\\ Higashi-Hiroshima 739-8526, Japan}
\affiliation[e]{Department of Mathematics, Birla Institute of Technology and
Science-Pilani,\\ Hyderabad Campus, Hyderabad-500078, India.}

\emailAdd{azzam-alwan@hiroshima-u.ac.jp}
\emailAdd{inagaki@hiroshima-u.ac.jp}
\emailAdd{bivu@hyderabad.bits-pilani.ac.in}
\emailAdd{shubhamn2616@gmail.com}

\abstract{Assuming static and spherically symmetric stars with perfect fluid matter, we used realistic equations of state to study neutron stars in covariant $f(Q)$ gravity. The structure profiles and properties of neutron stars such as mass, radius and compactness are obtained through numerical methods using quadratic, exponential, and logarithmic $f(Q)$ models. The results indicate that nonmetricity affects the interior profile deviations of the star, which in turn influence the properties of stars, as illustrated in the mass-radius relation diagram. This effect allows the star to accommodate either more or less matter compared to GR, resulting in a different total mass. For the quadratic model, we cannot generate larger masses, whereas the other two models can give consistent results for both smaller and larger masses of the observed stars. By tuning model parameters, we obtain $\mathcal{M}-\mathcal{R}$ diagrams that are compatible with observational constraints from NICER and LIGO.}


\begin{document}
\maketitle
\flushbottom

\section{Introduction}\label{sec:intro}
The geometrical modification of General Relativity (GR) has become inevitable post supernovae observations and other cosmological observations \cite{Riess1998, Perlmutter1999, Ade2016, Aghanim2016}. Extending and modifying GR presents a promising approach for addressing issues at both early and late cosmological epochs. GR traditionally employs Riemannian geometry, specifying the affine connection on the spacetime manifold to be metric compatible, specifically the Levi-Civita connection. However, different choices of affine connections on a manifold can lead to distinct but equivalent descriptions of gravity, potentially offering new insights \cite{Jimenez2019, Harada2020}. The Levi-Civita connection chosen by GR imposes that except curvature $R$, the other two fundamental geometrical objects, the nonmetricity $Q$ and torsion $T$, should both vanish. By relaxing these constraints, one can develop theories of gravity based on non-Riemannian geometry where curvature, torsion, and nonmetricity do not all necessarily vanish. For instance, by selecting a connection where both curvature and nonmetricity vanish while allowing torsion to be non-zero, one can formulate the Teleparallel Equivalent of GR (TEGR) \cite{Aldrovandi2013, Maluf2013}. Alternatively, a flat spacetime manifold with non-vanishing nonmetricity but no torsion leads to the Symmetric Teleparallel formulation of GR (STGR) \cite{Nester1998, Adak2005, Adak2008, Mol2014, Jarv2018, Jimenez2018, Jimenez2018a, Gakis2020}, in which a nonmetricity $Q$ mediates gravitational interaction.

The symmetric teleparallel gravity has evolved into coincident gravity or $f(Q)$ gravity [Jimenez et al \cite{Jimenez2018}]. As extended gravity theories have been emphasized in modern cosmology, alternative geometries are also being investigated. There are several works done in this gravity pertaining to cosmological implications in recent years, Ref. \cite{Harko2018,Lazkoz2019, Lu2019, Jimenez2020, Barros2020, Frusciante2021, Anagnostopoulos2021, Khyllep2021, Narawade2022, Narawade2023a, Narawade2023b, Heisenberg:2023lru,Heisenberg:2023wgk,Nojiri:2024zab}. In all of these studies, coincident gauges and line elements were used in Cartesian coordinates. Because of this specific choice, the covariant derivative is reduced to a partial derivative, simplifying calculations. However, the equations for pressure and energy are identical to those for $f(T)$. Regardless of whether the Universe is flat \cite{Subramaniam2023, Shabani2023, Paliathanasis2023} or curved \cite{Dimakis2022, Heisenberg2023, Shabani2024, Subramaniam2023a}, the $f(Q)$ theory with no coincident gauge attracts increased attention. As the Friedmann equations have been modified \cite{Dimakis2022}, the new gauge choices could affect the cosmological dynamics. As a result of the $f(Q)$ theory, the flat Universe always evolves in an unstable radiation era followed by a matter era and then a stable de-Sitter phase with a nontrivial affine connection \cite{Shabani2023}. By providing dark energy candidates, the $f(Q)$ theory can alleviate the cosmological constant problem. An affine connection could result in early-time acceleration following the inflationary period of the Universe \cite{Paliathanasis2023}. When taking into account non-zero spatial curvature, the early Universe might undergo a curvature-dominated phase. Furthermore, an open Universe might exhibit a peak in curvature density during intermediate times, according to \cite{Shabani2024}.

Besides its applications in cosmology, $f(Q)$ gravity has also been applied to compact astrophysical objects \cite{Bhar2023, Bhar2024, Kaur2024, Gul2024}. In this paper, we will discuss one of the compact objects, neutron stars (NS). These relativistic stars, which can be described by GR, serve as natural laboratories for studying high-density nuclear matter. Due to their extreme densities, strong gravitational fields, and the abundance of observational data that can be obtained, such as massive pulsar from Neutron star Interior Composition Explorer (NICER) \cite{Miller2019, Miller2021, Riley2021} and gravitational wave (GW) events from colaboration of Laser Interferometer Gravitational-Wave Observatory (LIGO) and Virgo Gravitational Wave Interferometer (Virgo) \cite{Abbott2017, Abbott2020}, neutron stars provide a good testing ground for both GR and modified gravity theories. The equation of state (EoS) for nuclear matter, which describes the relationship between density, pressure, and temperature, is key to understanding their properties and behavior \cite{Lattimer2016, Hebeler2013, Ozel2016, Steiner2017}. By examining the mass-radius relations, tidal deformations, and rotational dynamics of neutron stars \cite{Bertotti1969, Steiner2013, Steiner2014}, we can explore the validity of these theories under conditions that are inaccessible in the laboratory.

In addition to serving as testing grounds, modified gravity theories also help describe observational evidence that cannot be fully explained by GR. Several astrophysical observations have confirmed the existence of binary systems with NS having mass values that violate the Chandrasekhar limit for non-rotating degenerate stars to maintain stability \cite{Chandrasekhar1931}, indicating that neutron stars can possess significantly larger masses than previously predicted \cite{ Rawls2011, Mullally2009, Demorest2010, Zhang2019}. For example, modified gravity theories such as $f(R)$ gravity \cite{Ganguly2014, Astashenok:2014nua, Yazadjiev2014, Capozziello:2015yza, Astashenok:2017dpo} and $f(T)$ gravity \cite{Kpadonou2016, Pace2017, Fortes2022, Araujo2022} , can accommodate larger neutron star masses than GR, making them more flexible in meeting various observational constraints. In contrast to the other two trinity theories of gravity, there are still relatively few studies of neutron stars in $f(Q)$ gravity such as \cite{Lin2021}. They have used the polytropic EoS to calculate the structure and mass-radius relation for $f(Q) = Q + \alpha Q^2$ model. This study shows that $f(Q)$ gravity can accommodate massive star until $> 3 M_\odot$. In this work, we are motivate to calculated the profile of NS that includes the interior and exterior solutions of the stars. Also, we will obtain the properties of along with the compactness of the stars with $f(Q)$ covariant formulation. 

The paper is organized as: Section \ref{Sec:II} gives the covariant formalism of $f(Q)$ gravity and the derivation of gravitational field equations along with the TOV equation for $f(Q)$ gravity using the spherically symmetric metric. Section \ref{Sec:III} explains the forms of the $f(Q)$ models that we study. The mass-radius relation and the neutron star structure using the numerical solution for the $f(Q)$ models are analysed in the section \ref{Sec:IV}. In section \ref{Sec:V}, we discuss about the role of the nonmetricity $Q$ in the formation of the neutron stars. Also this section dedicated to the shortcomings occurring to form the neutron star using the considered $f(Q)$ models. The conclusions are presented in section \ref{Sec:VI}.


\section{Mathematical Fomalism}\label{Sec:II}
\subsection{Covariant Formulation of $f(Q)$ Theory}
The general affine connection can be decomposed into Levi-Civita connection $\left(\left\{_{~~\mu \nu}^\lambda\right\}\right)$, contortion $\left(K_{~~\mu \nu}^\lambda\right)$ and disformation $\left( L_{~~\mu v}^\lambda\right)$ as, 
    \begin{equation}\label{eq1}
    \Gamma_{~~\mu \nu}^\lambda= \left\{_{~~\mu \nu}^\lambda \right\}+K_{~~\mu \nu}^\lambda+L_{~~\mu \nu}^\lambda,
    \end{equation}
where
\begin{eqnarray*}
    \left\{_{~~\mu \nu}^\lambda \right\} = \frac{1}{2} g^{\lambda \alpha}\left(\partial_\mu g_{\alpha \nu}+\partial_\nu g_{\alpha \mu}-\partial_\alpha   g_{\mu \nu}\right),&&\quad
    K_{~~\mu \nu}^\lambda = \frac{1}{2}\left(T_{~~\mu\nu}^\lambda+T_{\mu~~\nu}^{~~\lambda}+T_{\nu~~\mu}^{~~\lambda}\right),\\
    L_{\ \mu \nu }^{\lambda} &=& \frac{1}{2}(Q_{~~\mu\nu }^{\lambda}-Q_{\mu~~\nu }^{\ \lambda}-Q_{\nu~~\mu }^{\ \lambda}).
\end{eqnarray*}
The torsion tensor $\mathcal{T}^{\lambda}_{~~\mu\nu}$, as well as the nonmetricity tensor $Q_{\lambda\mu\nu}$, are respectively presented for a spacetime equipped with the metric tensor $g_{\mu\nu}$ and the affine connection $\Gamma^{\lambda}_{~~\mu\nu}$.
\begin{eqnarray}\label{eq2}
    \mathcal{T}^{\lambda}_{~~\mu\nu} := \Gamma^{\lambda}_{~~\mu\nu}-\Gamma^{\lambda}_{~~\nu\mu},\quad\quad
    Q_{\lambda\mu\nu} := \nabla_{\lambda}g_{\mu\nu} = \partial_{\lambda}g_{\mu\nu}-\Gamma^{\alpha}_{~~\lambda\mu}g_{\alpha\nu}-\Gamma^{\alpha}_{~~\lambda\nu}g_{\alpha\mu}.
\end{eqnarray}
The nonmetricity scalar is defined as $Q = Q_{\lambda\mu\nu}P^{\lambda\mu\nu}$. The $P^{\lambda}_{~~\mu\nu}$ is called as nonmetricity conjugate and given as 
\begin{equation}\label{eq3}
   P^{\lambda}_{~~\mu\nu} = -\frac{1}{4}Q^{\lambda}_{~\mu \nu} + \frac{1}{4}\left(Q^{~\lambda}_{\mu~\nu} + Q^{~\lambda}_{\nu~~\mu}\right) + \frac{1}{4}Q^{\lambda}g_{\mu \nu}- \frac{1}{8}\left(2 \tilde{Q}^{\lambda}g_{\mu \nu} + {\delta^{\lambda}_{\mu}Q_{\nu} + \delta^{\lambda}_{\nu}Q_{\mu}} \right).
\end{equation}
A different definition of $Q = -Q_{\lambda\mu\nu}P^{\lambda\mu\nu}$ has been proposed in the literature, which changes the sign of the nonmetricity scalar $Q$. This is important to consider while comparing different $f(Q)$ results. STEGR (symmetric teleparallel equivalent) of GR can be produced if this nonmetricity scalar $Q$ replaces the Ricci scalar $R$ in the Einstein-Hilbert action. Because symmetric teleparallel theory is equivalent to GR, it inherits the same `dark' problem as general GR, so a modified $f(Q)$ gravity was introduced, in a similar manner to extending GR through a modified $f(R)$ theory. The components of the connection in eq. \eqref{eq1} can be rewritten as,
\begin{equation}\label{eq4}
\Gamma^{\lambda}\,_{\mu \beta }=\frac{\partial y^{\lambda}}{\partial \xi^{\rho }}\partial _{\mu }\partial _{\beta }\xi ^{\rho }.
\end{equation}
In the above equation, $\xi ^{\lambda}=\xi ^{\lambda}(y^{\mu })$ is an invertible relation and $\frac{\partial y^{\lambda}}{\partial \xi ^{\rho }}$ is the inverse of the corresponding Jacobian \cite{Jimenez2020}. This situation is called a coincident gauge, where there is always a possibility of getting a coordinate system with connections $\Gamma_{\ \mu \nu }^{\lambda}$ equaling zero. Hence, in this choice, the covariant derivative $\nabla _{\lambda}$ reduces to the partial derivative $\partial _{\lambda}$ i.e. $Q_{\lambda \mu \nu }=\partial _{\lambda}g_{\mu \nu }$. Thus, it is clear that the Levi-Civita connection $\left(\left\{_{~~\mu \nu}^\lambda\right\}\right)$ can be written in terms of the disformation tensor $L_{\ \mu \nu }^{\alpha }$ as $\left(\left\{_{~~\mu \nu}^\lambda\right\}\right)=-L_{\ \mu \nu }^{\lambda}$.
By varying the action term \cite{Jimenez2018, Zhao2022}
\begin{equation}\label{eq5}
    S = \int \frac{1}{2\kappa}f(Q)\sqrt{-g}~d^{4}x + \int \mathcal{L}_{m}\sqrt{-g}~d^{4}x,
\end{equation}
with respect to the metric tensor, we can obtain the field equation
\begin{equation}\label{eq6}
    \frac{2}{\sqrt{-g}}\nabla_{\lambda}\left(\sqrt{-g}f_{Q}P^{\lambda}_{~~\mu\nu}\right) - \frac{1}{2}g_{\mu \nu}f + f_{Q}(P_{\mu\lambda\alpha}Q^{~~\lambda \alpha}_{\nu} - 2Q_{\lambda \alpha \mu}P^{\lambda \alpha}_{~~~\nu}) = \kappa \mathcal{T}_{\mu \nu}~.
\end{equation}
Using this field equation, the covariant formulation has been developed and used effectively in studying geodesic deviations and cosmological phenomena \cite{Lin2021, Zhao2022, Beh2022, Subramaniam2023},
\begin{equation}\label{eq7}
    f_{Q}\mathring{G}_{\mu\nu}+\frac{1}{2}g_{\mu\nu}(Qf_{Q}-f)+2f_{QQ}P^{\lambda}_{~~\mu\nu}\mathring{\nabla}_{\lambda}Q = \kappa \mathcal{T}_{\mu \nu}~,
\end{equation}
where, $f_{Q}$ is derivative of $f$ with respect to $Q$ and $\mathring{G}_{\mu\nu} = R_{\mu\nu}-\frac{1}{2}g_{\mu\nu}R$, with $R_{\mu\nu}$ and $R$ are the Riemannian Ricci tensor and scalar respectively which are constructed by the Levi-Civita connection. For a linear form of $f(Q)$ function, the above equation reduces to GR. Variation of eq. \eqref{eq4} with respect to the connection, we can derive the equation of motion for the nonmetricity scalar as,
\begin{equation}\label{eq8}
    \nabla_{\mu}\nabla_{\nu}\left(\sqrt{-g}f_{Q}P^{\mu\nu}_{~~~\lambda}\right)=0.
\end{equation}

\subsection{TOV Equations in Covariant $f(Q)$ gravity}
Here, we are taking non-coincident gauge i.e. $\Gamma_{\ \mu \nu}^{\lambda}\neq 0$ into account and considering the spherically symmetric metric form as,
\begin{equation}\label{eq9}
    ds^{2} = -e^{A(r)}dt^{2}+e^{B(r)}dr^{2}+r^{2}(d\theta^{2}+sin^{2}\theta d\phi^{2}),
\end{equation}
with perfect fluid matter with $T_{\mu\nu}=diag\{-\rho c^2, p,p,p\}$ as ideal energy-momentum tensor. Using the Levi-Civita connection eq. \eqref{eq1} and the assumption of arbitrary affine connections, we can obtain all non-vanishing components of connections as,
\begin{eqnarray}\label{eq10}
 \Gamma^\theta_{~r\theta} &=& \Gamma^\theta_{~\theta r} = \Gamma^\phi_{~r\phi} = \Gamma^\phi_{~\phi r} = \frac{1}{r}, \quad \Gamma^{r}_{~\theta\theta} = -r\nonumber\\
 \Gamma^\phi_{~\theta\phi} &=& \Gamma^\phi_{~\phi\theta} = \cot\theta, \quad \Gamma^r_{~\phi\phi} = -r\sin^2\theta, \quad \Gamma^{\theta}_{~\phi\phi} = -\cos\theta\sin\theta.
\end{eqnarray}
We take this affine connection eq. \eqref{eq10} as the suitable affine connection for static spherically symmetric spacetime in $f(Q)$ theory and hence the equation of motion \eqref{eq7} becomes \cite{Zhao2022, Lin2021},
\begin{eqnarray}\label{eq11}
    \kappa\mathcal{T}_{tt} &=& \frac{e^{A-B}}{2r^{2}}\left\{r^{2}e^{B}f+2f_{Q}'r(e^{B}-1)+f_{Q}\left[(e^{B}-1)(2+rA')+(1+e^{B})rB'\right]\right\},\nonumber\\
    \kappa\mathcal{T}_{rr} &=& \frac{-1}{2r^{2}}\left\{r^{2}e^{B}f+2f_{Q}'r(e^{B}-1)+f_{Q}\left[(e^{B}-1)(2+rA'+rB')-2rA'\right]\right\},\nonumber\\
      \kappa\mathcal{T}_{\theta\theta} &=& -\frac{r}{4e^{B}}\left\{f_{Q}\left[-4A'-r(A')^{2}-2rA''+rA'B'+2e^{B}(A'+B')\right]+2e^{B}rf-2f_{Q}'rA'\right\},\nonumber\\
\end{eqnarray}
where $Q = \frac{(e^{-B}-1)(A'+B')}{r}$ and $f_{Q}' = f_{QQ}\frac{dQ}{dr}$. By solving eq. \eqref{eq11} with a concrete $f(Q)$ form and boundary conditions, we can obtain configurations $A(r)$ and $B(r)$. For example, if we consider vacuum solutions, this is $\mathcal{T}_{\mu\nu}=0$, then eq. \eqref{eq11} gives us $A'(r)+B'(r) = 0$. Eq. \eqref{eq11} can be expressed as a set of equations containing the Tolman-Oppenheimer-Volkov equations for $f(Q)$ gravity, which describe the structure of neutron stars, along with the continuity equation given by the energy-momentum conservation of $\mathcal{T}_{\mu\nu}$ as,
\begin{eqnarray}\label{eq12}
A'' &=& \frac{2 e^B \left( r (f(Q) + 2 p \kappa) + 
f_Q \left( A' + B' \right) \right)-A' \left( f_Q \left( 4 + r A' - r B' \right) + 
2 f_{QQ} r Q' \right)}{2f_Q r},\nonumber\\
B' &=& \frac{-\kappa e^B (p + \rho) r + f_Q A'}{f_Q}, \nonumber\\
p' &=& -\frac{(p+\rho)}{2}A'.
\end{eqnarray}
It is important to note that the conservation of energy-momentum in $f(Q)$ gravity with the spherical symmetric metric form remains an issue \cite{Zhao2022, De2023}. This issue is further discussed in \hyperref[AppendixA]{Appendix A}. By using the EoS and providing initial values for $A$, $B$, $Q$, $\rho$, and $p$, we can now describe the structure of neutron stars in $f(Q)$ gravity. If we set $f(Q)=Q$ in eq. \eqref{eq12}, we can easily obtain the GR case solution (\hyperref[AppendixB]{see Appendix B}). From these equations, we can calculate the structure of neutron stars using various models.


\section{$f(Q)$ Models}\label{Sec:III}
In the study of neutron star structure within the framework of $f(Q)$ gravity, we consider quadratic, exponential and logarithmic $f(Q)$ models. These models modify the gravitational action by introducing nonlinear functions of the nonmetricity scalar $Q$, thereby altering the equations governing the stellar structure. By analyzing these models, we aim to understand the impact of such modifications on the properties of neutron stars. 
\subsection{$f(Q)=Q + \alpha Q^2$}
To explore the effects of different $f(Q)$ models, we first consider the specific model given by \cite{Harko2018}:
\begin{equation}\label{eq13}
f(Q)=Q+ \alpha Q^2,
\end{equation}
where $\alpha$ is the parameter for the quadratic nonmetricity correction. This specific choice is the simplest one and it is inspired in the Starobinsky model in $f(R)$ gravity, which has this same functional form. This quadratic term is particularly suitable for neutron star cases that exhibit strong gravity regimes. Conversely, the linear term typically applies to low gravity regimes. Additionally, this model is a special case of the more general power-law form of $f(Q)$, where:
\begin{equation}\label{eq14}
f(Q)=Q+ \alpha Q^n,
\end{equation}
with $n=2$. While the power-law form with different values of $n$ could offer a broader range of behaviors and insights, numerical difficulties were encountered when working with values other than $n=2$. Therefore, for the purposes of this paper, we focus on the $Q^2$ case. This choice allows for a more manageable numerical treatment while still capturing essential aspects of the modifications to gravity in the strong field regime relevant to neutron stars. With subtitute eq. \eqref{eq13} to eq. \eqref{eq12}, we can get the TOV equations as,
\begin{eqnarray}\label{eq15}
A''  &=& \frac{2 e^B r (Q + Q^2 \alpha + 2 p \kappa) + 2 e^B (1 + 2 \alpha) \left( A' + B' \right)}{2 (r + 2 r \alpha)} \nonumber\\
&&+ \frac{A' \left( - \left( (1 + 2 \alpha) (4 + r A' - r B') \right) - 4 r \alpha Q' \right)}{2 (r + 2 r \alpha)},\nonumber\\
B' &=& \frac{e^B r \kappa (p + \rho)}{1 + 2 Q \alpha} - A'.
\end{eqnarray}
We can easily revert the eq. \eqref{eq14} to the GR case by setting the parameter $\alpha=0$. Previous study \cite{Lin2021} have discussed this model, demonstrating good consistency in both interior and exterior solutions using a polytropic equation. Moreover, the parameter $\alpha$ has an important role in generating the mass of neutron stars, where positive values of $\alpha$ yield smaller masses and negative values yield larger masses. In this paper, we re-examine the $Q^2$ model using more realistic EoS.

\subsection{$f(Q)=Q + \alpha e^{\beta Q}$}
Having derived the quadratic model, we now turn our calculation to another $f(Q)$ model, specifically the exponential form given by \cite{Anagnostopoulos2021}:
\begin{equation}\label{eq16}
    f(Q) = Q + \alpha e^{\beta Q},
\end{equation}
where $\beta$ and $\alpha$ are the parameters for the exponential correction term. According to \cite{Anagnostopoulos2021}, the exponential model provides slightly better fits to cosmological data than the concordance model. Also the exponential function allows us to explore more complex non-linear effects in strong gravitational fields, such as those found in neutron stars. Additionally, this model is connected to scalar-tensor theories and has been used in cosmology or large scale structure \cite{Sokoliuk2023, Narawade2023b}. Here we little modify the model for avoiding numerical problem from our TOV. The exponential model can capture details in areas with strong gravity that linear or quadratic models might miss. Moreover, the parameters $\alpha$ and $\beta$ can be adjusted to better match observations. Specifically, $\beta$ controls the rate of exponential growth, allowing for precise adjustments, while $\alpha$ scales the overall amplitude of the exponential correction, providing broader modifications to the model. These parameters help in achieving a stable neutron star solution that fits with observational constraints. Because of its versatility feature, the exponential model $ f(Q) = Q + \alpha e^{\beta Q} $ is a good choice for studying the structure of neutron stars. 
Using eq. \eqref{eq12}, we can derive the TOV equations for this model as,  
\begin{eqnarray}\label{eqexp}
A'' &=& \frac{A' \left( -4 + 2 e^B + r B' + e^{Q \beta} \alpha \beta \left( -4 + 2 e^B + r B' - 2 r \beta Q' \right) \right)}{2 \left( r + e^{Q \beta} r \alpha \beta \right)} \nonumber\\
&&+\frac{2 e^B \left( Q r + 2 p r \kappa + B' + e^{Q \beta} \alpha \left( r + \beta B' \right) \right) - A'^2 r \left( 1 + e^{Q \beta} \alpha \beta \right)}{2 \left( r + e^{Q \beta} r \alpha \beta \right)}, \nonumber\\
B' &=& \frac{e^B \kappa r (p + \rho)}{1 + e^{Q \beta} \alpha} - A'.
\end{eqnarray} 
We can easily revert eq. \eqref{eq16} to the GR case by setting the parameters $\alpha=0$. This ensures that the solutions obtained are consistent with the standard GR in the absence of modifications.

\subsection{$f(Q)=Q - \alpha \ln{(1-\beta Q)}$}
In addition to the quadratic and exponential models, the logarithmic model is also considering. This model has been well-tested for explaining cosmological phenomena and dark energy \cite{Najera2023}. Here we consider
\begin{equation}\label{eq19}
    f(Q)=Q - \alpha \ln{(1-\beta Q)},
\end{equation}
Geometrically, logarithmic model was successful in predicting the cosmic late-time accelerated expansion, and it is also a strong candidate for solving the cosmological constant problem. Moreover this model is effective in compressing the correction terms, which allows for the creation of more stable neutron stars with higher masses compared to the quadratic model. Similar to the exponential model, the parameters $\alpha$ and $\beta$ have effects in adjusting the model to fit observational data accurately. Specifically, $\alpha$ affects the rate at which the logarithmic function approaches its critical point, and $\beta$ scales the impact of the logarithmic term. This flexibility benefits the logarithmic model in achieving stable and good solutions for the structure of neutron stars. From eq. \eqref{eq12}, we can derive the modified TOV as,
\begin{eqnarray}\label{eqlog}
A'' &=& \frac{e^B \alpha (-1 + Q \beta) \ln(1 - Q \beta)}{1 - Q \beta + \alpha \beta} + \frac{(-1 + Q \beta) (Q + 2 p \kappa)r + B'(-1 + Q \beta - \alpha \beta)}{(-1 + Q \beta - \alpha \beta)e^{-B}}-\frac{A'^2}{2}\nonumber\\
&&+ A' \left( \frac{-2 + e^B}{r} + \frac{B'}{2} - \frac{\alpha \beta^2 Q'}{(-1 + Q \beta) (-1 + Q \beta - \alpha \beta)} \right),\nonumber\\
B' &=& \frac{e^B \kappa r (-1 + Q \beta) (p + \rho)}{-1 + Q \beta - \alpha \beta} - A'.
\end{eqnarray}
By setting $\beta=0$, we can recover GR case for this model. In the following sections, we will analyze how these three models influences the structure of neutron stars and evaluate its compatibility with observational data using realistic EoS.


\section{Neutron Star Structure for $f(Q)$ Models}\label{Sec:IV}
Using the TOV equations derived for each model, we perform numerical calculations to obtain the interior and exterior profiles of the neutron stars. Through these calculations, we can see how nonmetricity affects the structure of the stars. Furthermore, we calculate the mass-radius relation for each model, comparing these results with observational constraints to test the consistency of the $f(Q)$ models.

\subsection{Boundary and Junction Conditions}
To solve our ODE systems for all the $f(Q)$ models, we have three equations:\\ $A''=f_1 (Q,Q',B,B',p,\rho,r)$, $B'=f_2 (Q,B,A',p,\rho, r)$, and $p'=f_3 (A',p,\rho)$. Unlike the GR case, where the term $Q'$ vanishes because $f_{QQ} = 0$, in all three models, we need an additional equation $Q'$ that can be derived from eq. $\eqref{eq11}$ to decouple the behavior of our ODE system. This allows us to set initial values for our ODE systems. In this discussion, we use the following initial values \cite{Lin2021}:
\begin{equation}\label{eq20}
B=0, \quad A'=0, \quad A=A_0, \quad Q=0, \quad p=p_c.
\end{equation}
The constant $A_0$ can be determined by matching the interior and exterior solutions of the star. As discussed in the covariant formulation of $f(Q)$, the condition $A' + B' = 0$ in vacuum implies that $Q(r) = 0$ outside the star. By evaluating eq. \eqref{eq11} under vacuum conditions, we can obtain:
\[ e^B (2 f_Q + f(Q) r^2) - 2 f_Q (1 + r A') = 0. \]
Given the relationship $A'(r) + B'(r) = 0$ and $e^{A(r)} = e^{-B(r)}$, which can be easily derived from eq. \eqref{eq11}, we can obtain the exterior solution of the star as:
\begin{equation}\label{eqext}
e^{-B(r)} = e^{A(r)} = 1 + \frac{C}{r} + \frac{f(Q)|_{0}}{6f_Q|_0}r^2,
\end{equation}
where $C$ is an integration constant. This solution is similar to the Schwarzschild-de Sitter (SdS) solution with the cosmological constant $\Lambda=\frac{f(Q)|_{0}}{2f_Q|_0}$. When we apply the three $f(Q)$ models, the models given by Eqns. \eqref{eq13} and \eqref{eq19} will result in the same exterior solution as in GR, because the third term in eq. \eqref{eqext} will vanish. However, in the case of eq. \eqref{eq15}, the third term does not vanish. A similar solution can also be obtained from eq.\eqref{eq7} by directly using the vacuum condition and $Q(r) = 0$. These solutions provide a boundary condition at the surface of the star that ensures the correct asymptotic behavior at infinity. The surface of the star can be determined by applying the boundary condition on the pressure. At the surface, the pressure of the star approaches zero, which can be defined as
\begin{equation}\label{eq22}
    p(r_s)\approx 0,
\end{equation}
where $r_s$ is the radius of the star at its surface. Another important aspect to consider is the junction condition for solving the equations between the interior and exterior solutions of these ODEs, addressing the transition. Since we are not considering any additional scalar fields, and as seen in eq.~\eqref{eq7}, which still contains a term proportional to the Einstein tensor $ G_{\mu\nu} $, we can apply the GR junction conditions. These conditions are given by $ [h_{\mu\nu}] = 0 $ and $ [K_{\mu\nu}] = 0 $ \cite{Israel1966, Marolf2005}. The notation $ [...] $ specifically indicates the jump or discontinuity across the surface. The condition $ [h_{\mu\nu}] = 0 $ implies that the induced metric must be continuous across the hypersurface. On the other hand, $ [K_{\mu\nu}] = 0 $ indicates that there should be no discontinuity in the extrinsic curvature across the hypersurface. This condition ensures that the embedding of the hypersurface in the spacetime does not introduce any physical inconsistencies or abrupt changes. If we look at other modified gravity models such as $f(R)$ \cite{Deruelle2008, Senovilla2013, Feng2017}, we also need to pay attention to the continuity of nonmetricity. In this case, since we only have $ Q' $ in our TOV equations, it is necessary that $ [Q] = 0 $ so we can assume that there is no delta-function-like discontinuity. From eq.~\eqref{eqext}, the exterior solution is identical to the GR case, except for the exponential model which will be discussed later. This allows us to use the same junction conditions as in GR. This condition is also applied in the study of $f(Q)$ gravity on other compact objects. For instance, in \cite{Maurya2022, Maurya2023}, the SdS solution is used as the exterior solution for strange star cases, and the same conditions apply to the interior and exterior boundaries. In other cases of charged isotropic compact stars, such as \cite{Chaudharya2024}, the RdS (Reissner-Nordstr{\"o}m-de Sitter) solution is used as the exterior solution, with a similar procedure applied for the junction condition between the interior and exterior solutions. In these cases, we only consider non-charged stars, so the RdS solution will have the same form as the SdS solution. For the mass of stars ($m$), we use explicitly the following equation refer to \cite{Lin2021}:
\begin{equation}\label{eq21}
m=4\pi \int^{r_s}_0 \rho(r)r^2 dr~.
\end{equation}

To solve the ODEs numerically, we also need to use dimensionless physical variables. We rescale using the gravitational radius $r_g$, so that the physical variables become,
\begin{equation}\label{eq23}
\hat{r}\rightarrow \frac{r}{r_g}, \quad \hat{p}\rightarrow \frac{p}{p_g}, \quad \hat{\rho} \rightarrow \frac{\rho}{\rho_g}, \quad \hat{Q}\rightarrow Q r_g^2, \quad \hat{m}\rightarrow \frac{m}{M_\odot},
\end{equation}
where
\begin{equation}\label{eq24}
r_g= \frac{G M_\odot}{c^2}, \quad p_g = \frac{M_\odot c^2}{r_g^3}, \quad \rho_g = \frac{M_\odot}{r_g^3}.
\end{equation}
In this context, $M_\odot$ represents the mass of the Sun, $c$ denotes the speed of light, and $G$ be the gravitational constant. All these constants are expressed in the cgs unit system, where $M_\odot\approx 1.989 \times 10^{33}$ $g$, $c\approx 2.997 \times 10^{10}$ $cm/s$, and $G\approx 6.674 \times 10^{-8}$ $dyne.cm^2/g^2$. Another problem arises from the singularity behavior in our TOV equations. To avoid this, we expand eq. \eqref{eq12} around $r=0$, obtaining the asymptotic solution near the center of the stars. This expansion helps in preventing singularities during the numerical integration. For the EoS, we used SLy \cite{Douchin2001, Potekhin2013} tabulation\footnote{Tabulated SLy EoS can be obtained from \url{http://www.ioffe.ru/astro/NSG/NSEOS/}.}. We also used the piece-wise polytrope form of APR4 \cite{APR4} and MS1b \cite{MS1b}, utilizing tabulation parameters from \cite{Read2009}. A fixed three-piece fit is implemented at $10^{14.7} \text{g/cm}^3$ and $10^{15} \text{g/cm}^3$ for both EoS. By using these EoS, we can achieve a more realistic modeling of the pressure-density relations ($p=p(\rho)$) within neutron stars. After doing the numerical setting, we can now solve the ODE systems. We use the RK45 method, which can dynamically adjust the step size to balance accuracy and computational efficiency, with the \texttt{scipy.integrate.solve\_ivp} package in Python to numerically integrate the ODEs and obtain the solutions. This package allows for flexible and efficient solving of initial value problems for ODEs. 

\subsection{Numerical Solutions}\label{subsec:4.2}
\begin{figure}[H]
    \centering
    \begin{subfigure}[b]{0.50\textwidth}
        \centering
        \includegraphics[width=\linewidth]{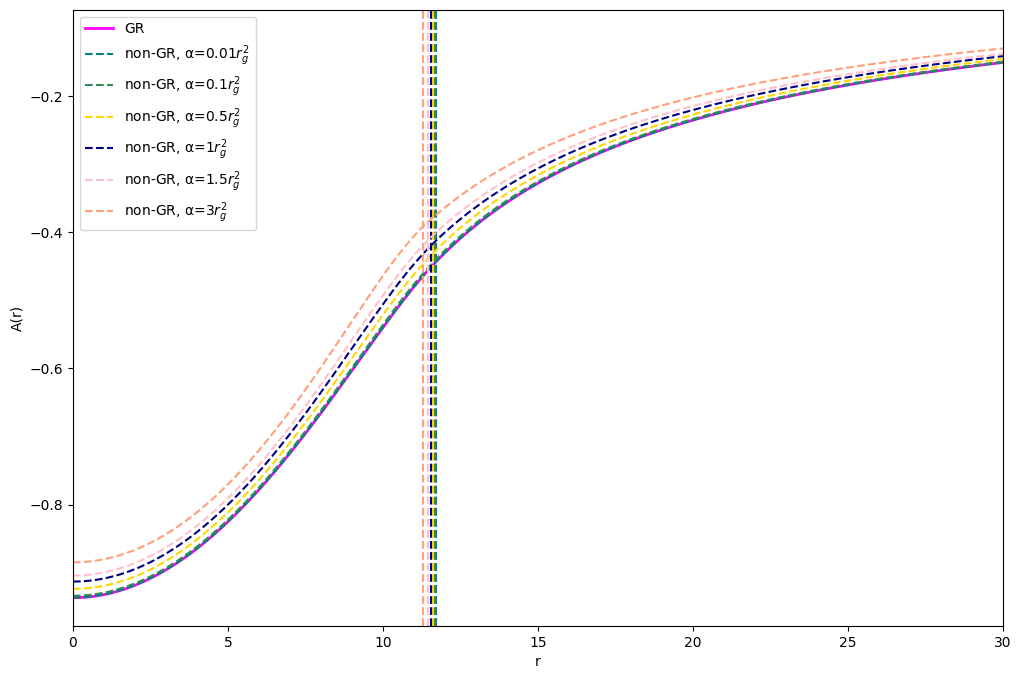}
        \caption{Metric Component of $A(r)$ for $f(Q)=Q+\alpha Q^2$}
        \label{fig:ASLy}
    \end{subfigure}%
    \hfill
    \begin{subfigure}[b]{0.48\textwidth}
        \centering
        \includegraphics[width=\linewidth]{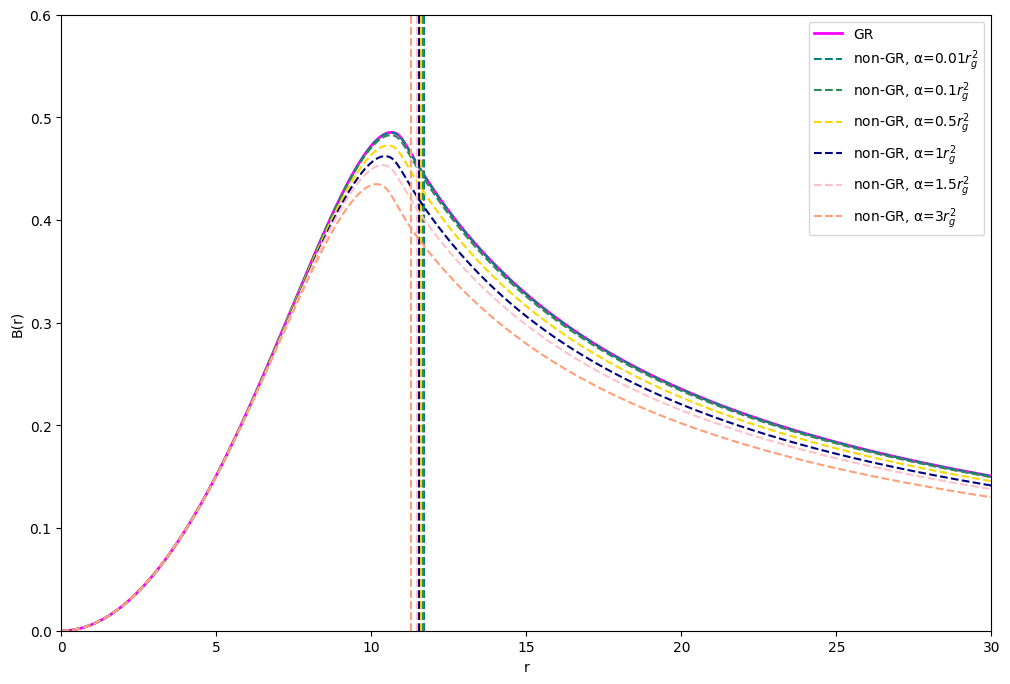}
        \caption{Metric Component of $B(r)$ for $f(Q)=Q+\alpha Q^2$}
        \label{fig:BSLyQ2}
    \end{subfigure}%
    \vskip\baselineskip
    \begin{subfigure}[b]{0.48\textwidth}
        \centering
        \includegraphics[width=\linewidth]{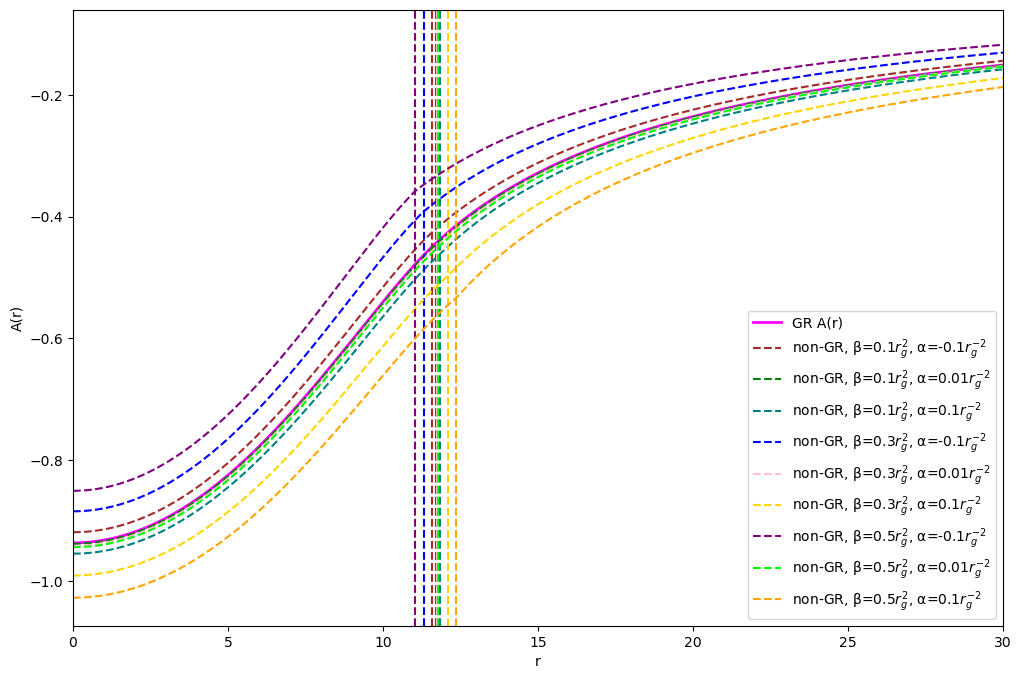}
        \caption{Metric Component of $A(r)$ for $f(Q)=Q+\alpha e^{\beta Q}$}
        \label{fig:ASLyQexp}
    \end{subfigure}
    \hfill
    \begin{subfigure}[b]{0.48\textwidth}
        \centering
        \includegraphics[width=\linewidth]{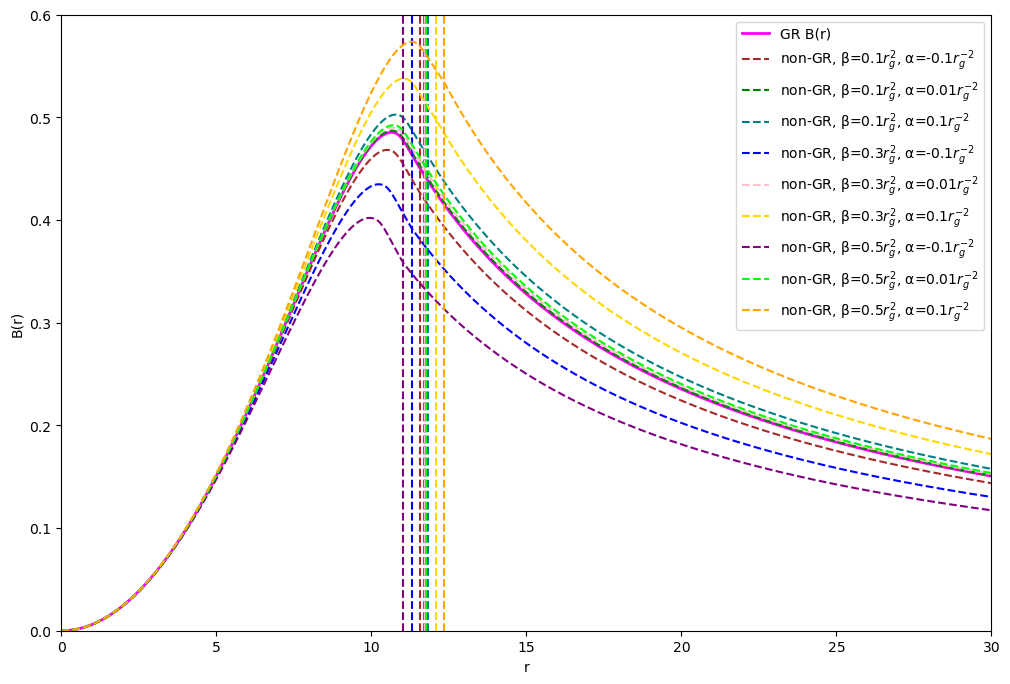}
        \caption{Metric Component of $B(r)$ for $f(Q)=Q+\alpha e^{\beta Q}$}
        \label{fig:BSLyQexp}
    \end{subfigure}
    \vskip\baselineskip
    \begin{subfigure}[b]{0.48\textwidth}
        \centering
        \includegraphics[width=\linewidth]{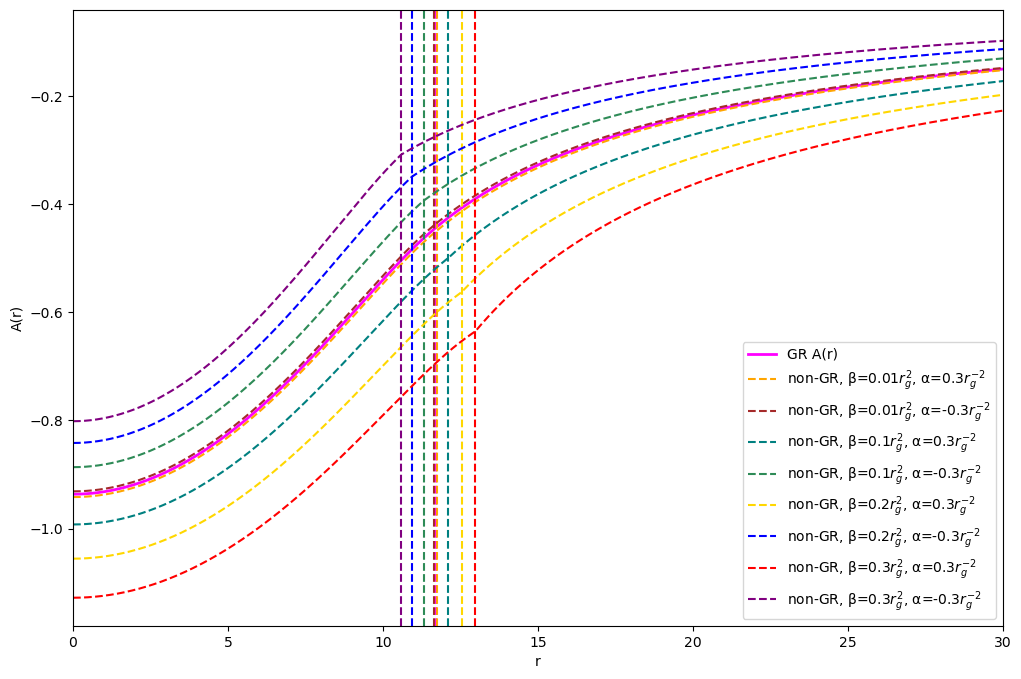}
        \caption{Metric Component of $A(r)$ for $f(Q)=Q-\alpha \ln(1-\beta Q)$}
        \label{fig:ASLyQlog}
    \end{subfigure}
    \hfill
    \begin{subfigure}[b]{0.48\textwidth}
        \centering
        \includegraphics[width=\linewidth]{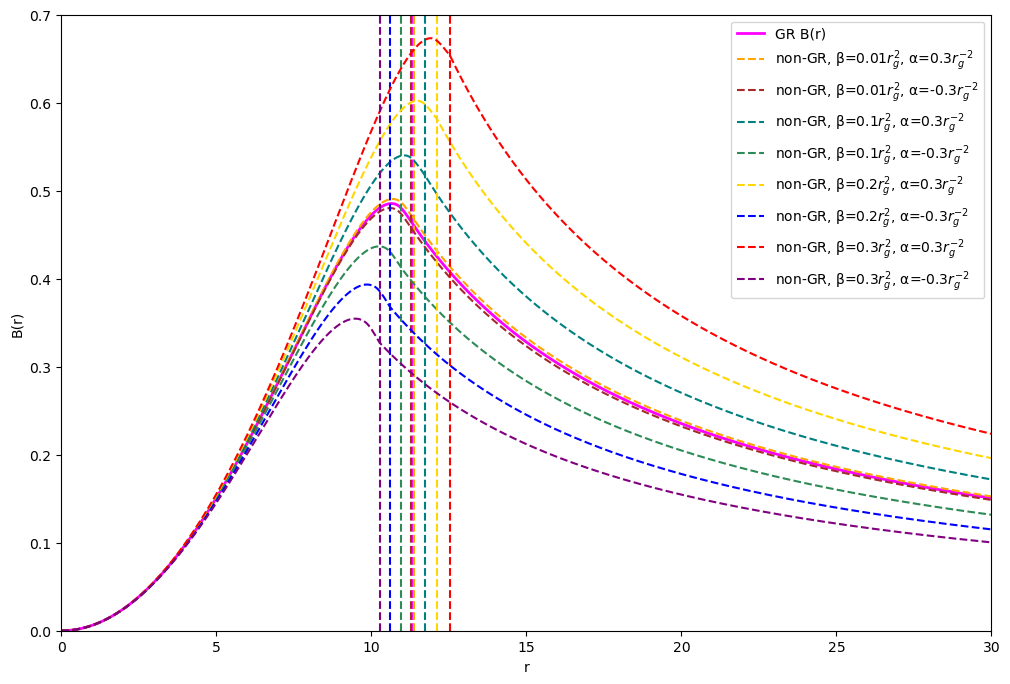}
        \caption{Metric Component of $B(r)$ for $f(Q)=Q-\alpha \ln(1-\beta Q)$}
        \label{fig:BSLyQlog}
    \end{subfigure}
    \caption{Metric solutions for the $f(Q)$ models with various parameters $\alpha$ and $\beta$ using $\rho_c=1\times10^{15}, \text{g/cm}^3$ (SLy EoS). $r$ is in $km$ unit. The parameters $\alpha$ and $\beta$ affect the deviation in each solution. The vertical dashed lines indicate the $r_{\text{surface}}$ for each set of parameters. We can see that for small values of $\alpha=0.01r_g^{2}$ and $\alpha=0.01r_g^{-2}$, the $Q^2$ and exponential models closely match the GR case. Similarly, for the logarithmic model, setting a very small parameter $\beta=0.01r_g^{-2}$ also allows it to match the GR case. The functions $A(r)$ and $B(r)$ are connected from the interior metric to the exterior metric using the SdS solution at the $r_s$.}
    \label{fig:ABSLy}
\end{figure}

\begin{figure}[H]
    \centering
    \begin{subfigure}[H]{0.50\textwidth}
        \centering
        \includegraphics[width=\linewidth]{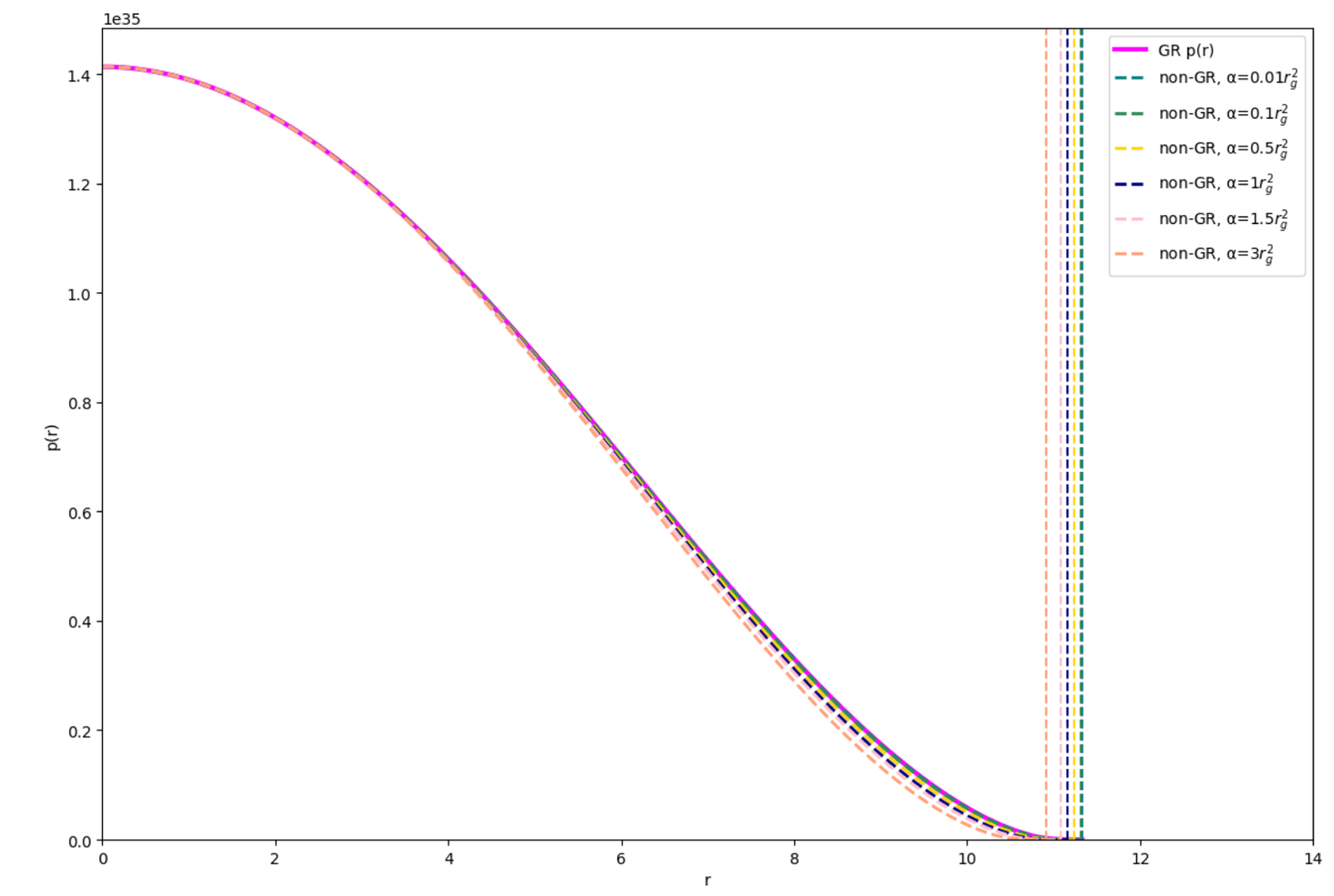}
        \caption{Pressure profile of $A(r)$ for $f(Q)=Q+\alpha Q^2$}
        \label{fig:pSLy}
    \end{subfigure}%
    \hfill
    \begin{subfigure}[H]{0.48\textwidth}
        \centering
        \includegraphics[width=\linewidth]{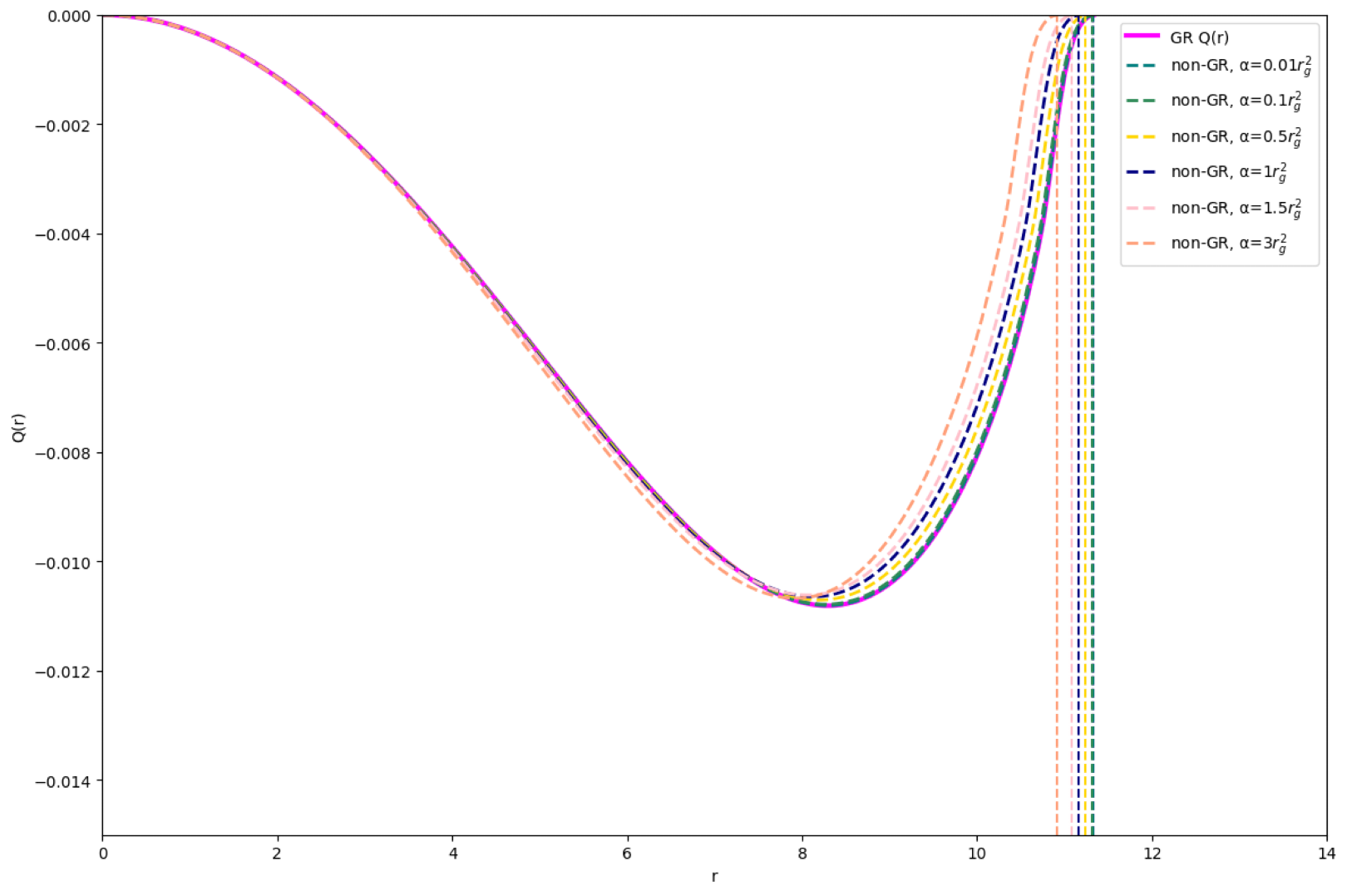}
        \caption{nonmetricity Profile of $Q(r)$ for $f(Q)=Q+\alpha Q^2$}
        \label{fig:QSLyQ2}
    \end{subfigure}%
    \vskip\baselineskip
    \begin{subfigure}[H]{0.48\textwidth}
        \centering
        \includegraphics[width=\linewidth]{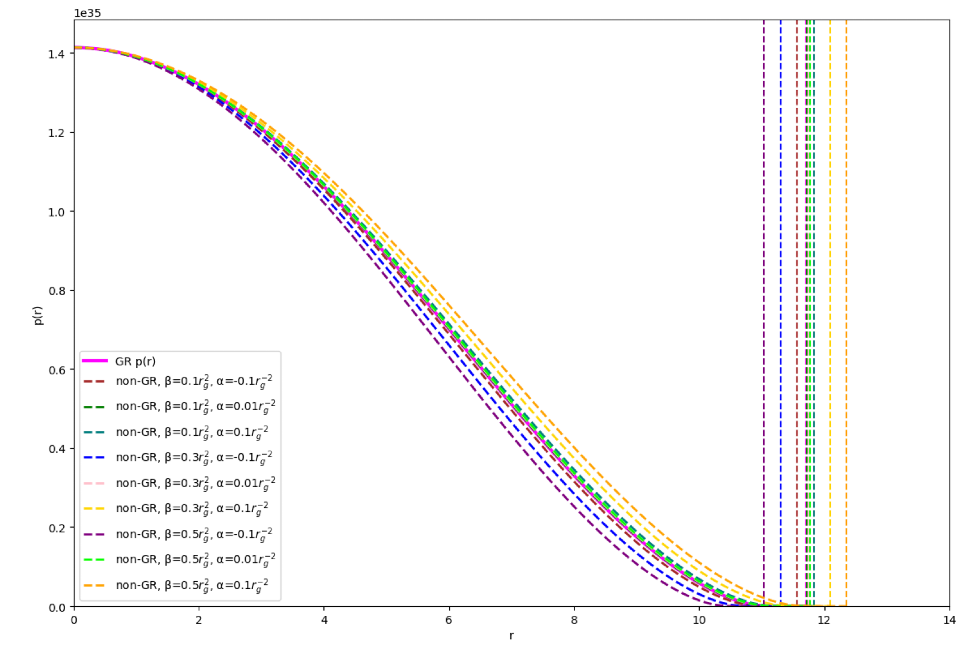}
        \caption{Pressure profile of $p(r)$ for $f(Q)=Q+\alpha e^{\beta Q}$}
        \label{fig:pSLyQexp}
    \end{subfigure}
    \hfill
    \begin{subfigure}[H]{0.48\textwidth}
        \centering
        \includegraphics[width=\linewidth]{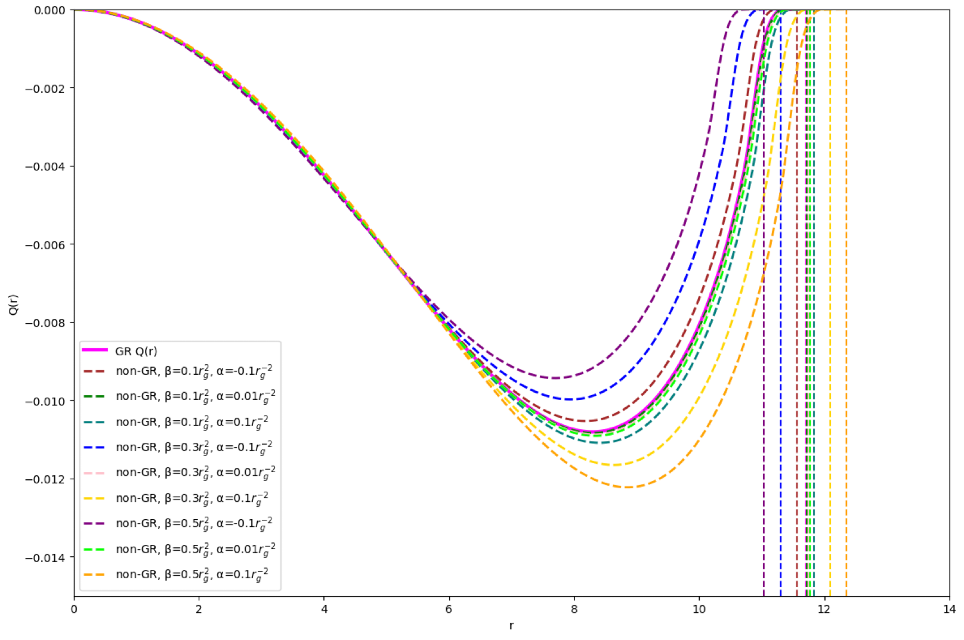}
        \caption{nonmetricity Profile of $Q(r)$ for $f(Q)=Q+\alpha e^{\beta Q}$}
        \label{fig:QSLyQexp}
    \end{subfigure}
    \vskip\baselineskip
    \begin{subfigure}[H]{0.48\textwidth}
        \centering
        \includegraphics[width=\linewidth]{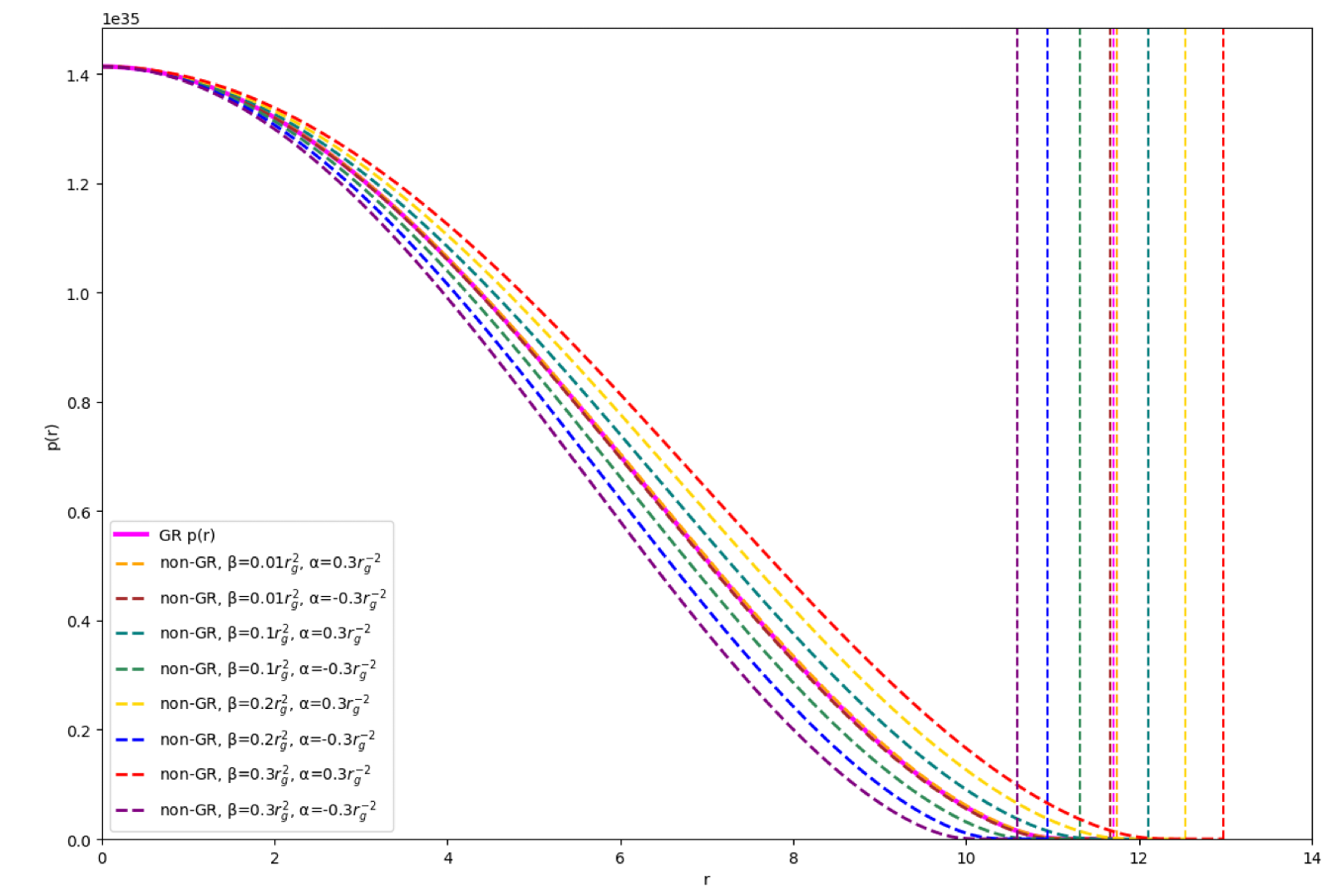}
        \caption{Pressure profile of $p(r)$ for $f(Q)=Q-\alpha \ln(1-\beta Q)$}
        \label{fig:pSLyQlog}
    \end{subfigure}
    \hfill
    \begin{subfigure}[H]{0.48\textwidth}
        \centering
        \includegraphics[width=\linewidth]{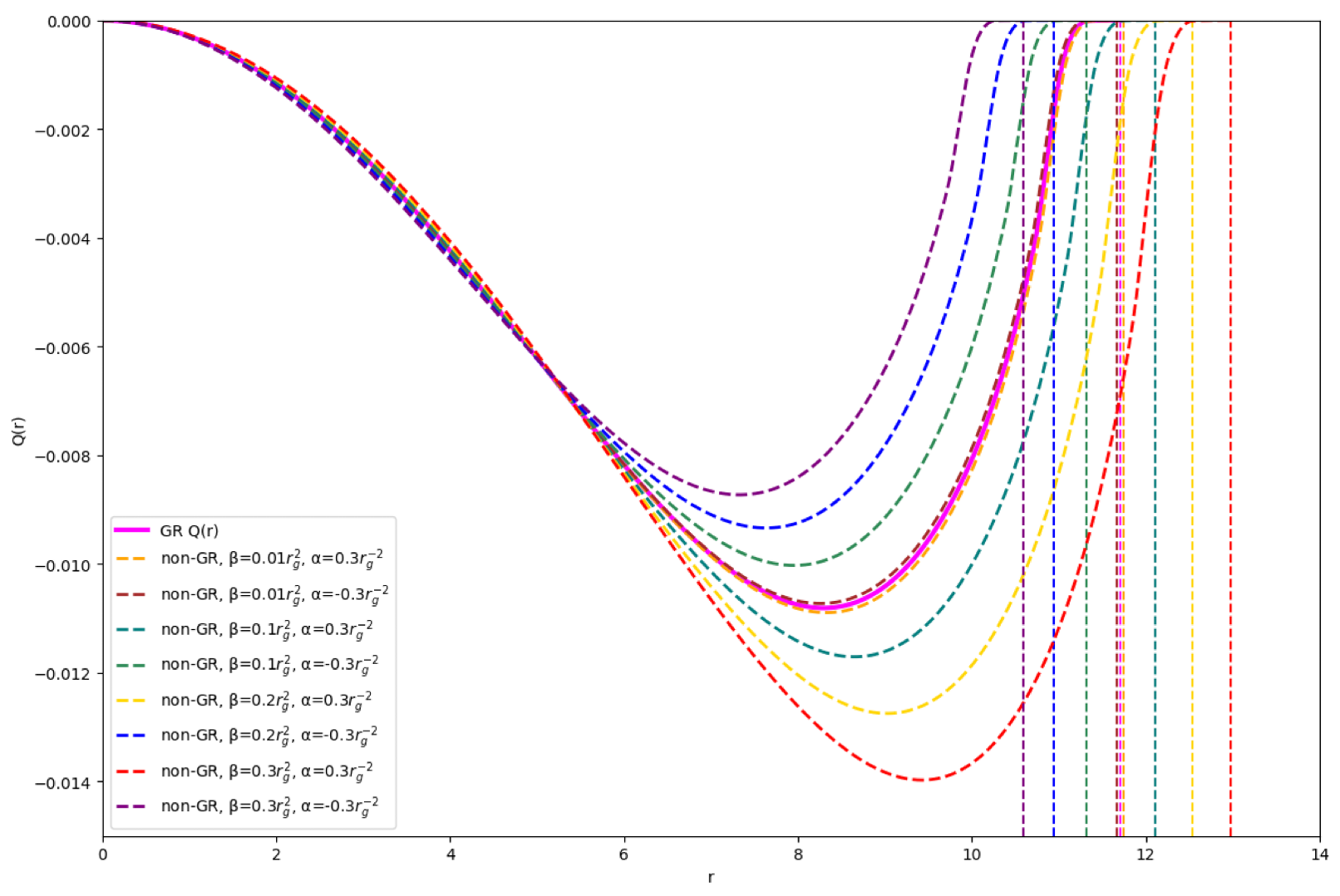}
        \caption{nonmetricity Profile of $Q(r)$ for $f(Q)=Q-\alpha \ln(1-\beta Q)$}
        \label{fig:QSLyQlog}
    \end{subfigure}
    \caption{Nonmetricity and pressure profile for the $f(Q)$ models with various parameters $\alpha$ and $\beta$ using $\rho_c=1\times10^{15}, \text{g/cm}^3$ (SLy EoS). $p(r)$ is in $dyne/cm^2$ unit, $Q(r)$ is in $r_g^{-2}$ unit, and $r$ is in $km$ unit. The vertical dashed lines indicate the $r_{\text{surface}}$ for each set of parameters. The profiles of $Q(r)$ and $p(r)$ approaching zero at the surface boundary match the exterior solutions of $A(r)$ and $B(r)$.}
    \label{fig:PQSLy}
\end{figure}

We compute the structure of neutron stars and the mass-radius diagrams for the nonmetricity formulation. Due to numerical limitations in our TOV equations, we will only calculate for positive $\alpha$ for the $Q^2$ model. The solutions of the TOV equations for all three models are illustrated in figure~\ref{fig:ABSLy} and figure \ref{fig:PQSLy}. As shown in these figures, the central pressure of the star rapidly drops to zero. This boundary condition allows us to determine the surface of the star ($r_s$) where $p(r_s) \approx 0$. Due to numerical reasons, we set $p(r_s) \leq 10^{-8} p_c$ for determining $r_s$. At this radius, we identify the boundary of the star and solve the junction conditions between the interior and exterior solutions of the star. In this study, we only consider SdS solution for the exterior of the stars. Using eq. \eqref{eqext} at the vacuum condition, we determine $C$ using the shooting method simultaneously with $A_0$. For $f(Q)=Q+\alpha Q^2$ and $f(Q)=Q-\alpha \ln(1-\beta Q)$, we can easily get $\Lambda=0$, but for $f(Q)=Q+\alpha e^{\beta Q}$, $\Lambda \neq 0$, which will make the exterior solution of the exponential model different. This is because $f(Q)|_{0}=\alpha$, so it is obtained that $\Lambda = \frac{\alpha}{2(1+\alpha \beta)}$. However, since $\alpha = \hat{\alpha}/r_g^2$ for the exponential model, the constant $r_g^2$ compresses the third term of the SdS solution, resulting in an exterior solution similar to the other two models as illustrated in figure \ref{fig:ABSLy}. This process ensures continuity for the metric functions $B(r)$ and $A(r)$ between interior and exterior solution. In the figures, we can see also that the exterior solutions $B(r)$ and $A(r)$ naturally converge to zero as $r \to \infty$ because $1+\frac{C}{r}$ exterior solution.

Next, we turn our attention to the behavior of the nonmetricity profile $Q(r)$, which is illustrated in figure~\ref{fig:QSLyQ2},~\ref{fig:QSLyQexp},~\ref{fig:QSLyQlog}. For $f(Q) = Q + \alpha Q^2$, the nonmetricity decreases as $\alpha$ increases. This is the opposite of the other two models, where a positive $\alpha$ results in a more dominant nonmetricity compared to the GR case. Specifically, in the $Q^2$ model, $Q^2$ is positive while $Q$ is negative. This solution aligns with the solutions for $B(r)$ and $A(r)$. By calculating $Q_{rrr} = e^B B'$, we observe that the value of $B(r)$ can decrease as $Q(r)$ decreases, suggesting that a larger $\alpha$ results in a smaller $B(r)$. On the other hand, the behavior of $A(r)$ is consistent with the solutions for $Q_{rtt}$, where $A(r)$ becomes smaller as $Q(r)$ decreases. This indicates that the nonmetricity has important role in the formation of the structure of neutron stars. This conclusion is strengthened by the neutron star mass-radius diagram which will be discussed further in the next subsection. This behavior is also consistent with previous study related to solutions for $A(r)$, $B(r)$, and $Q(r)$. However, there is a slight difference where previous studies found that $Q(r)$ values were larger for positive $\alpha$ in the $Q^2$ model. We expect this different result arises because we used all field equations to obtain the TOV equations. Our results of all three models show the consistent relationship between $Q(r)$, $A(r)$, and $B(r)$.

\newpage

\newpage
\subsection{Mass-Radius Relation}
\begin{figure}[H]
    \centering
    \begin{subfigure}{0.48\textwidth}
        \centering
        \includegraphics[width=\linewidth]{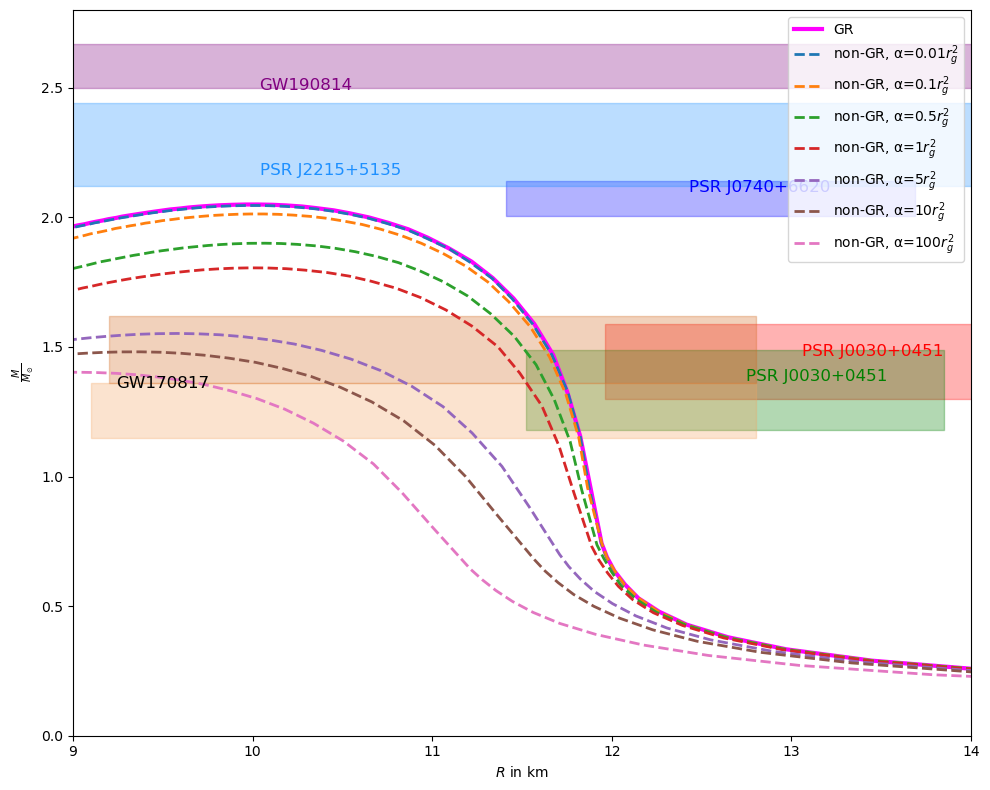}
        \caption{Mass-Radius Diagram for SLy EoS}
        \label{fig:MRSLy}
    \end{subfigure}%
    \hfill
    \begin{subfigure}{0.48\textwidth}
        \centering
        \includegraphics[width=\linewidth]{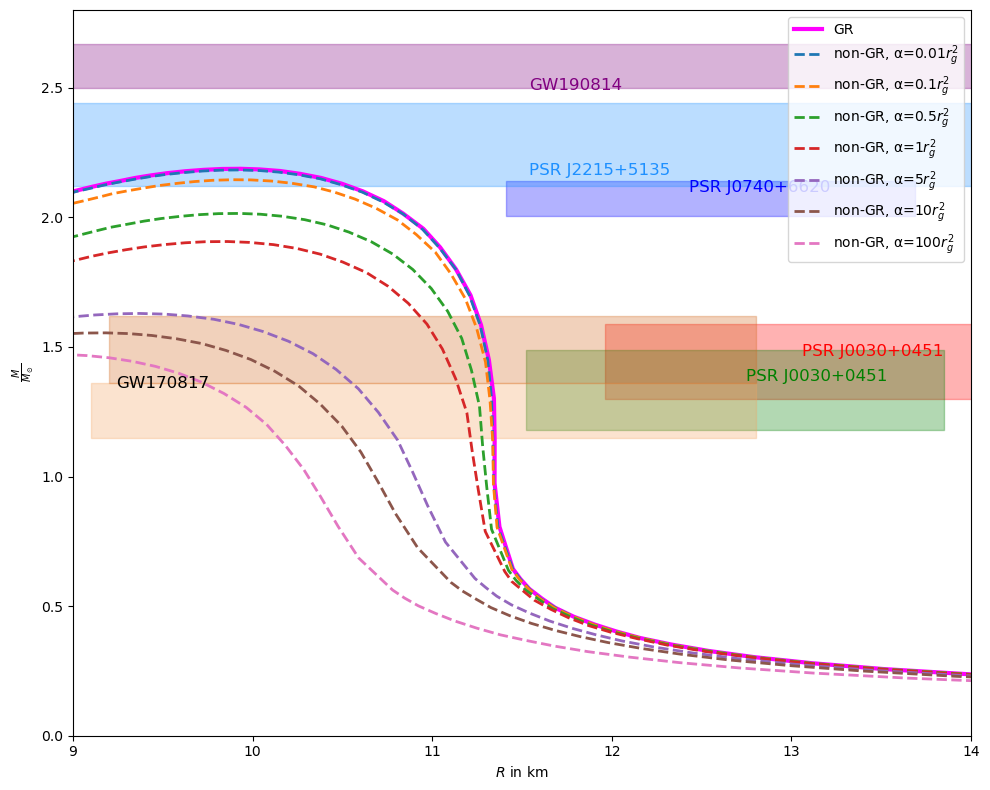}
        \caption{Mass-Radius Diagram for APR4 EoS}
        \label{fig:MRAPR4}
    \end{subfigure}%
    \vskip\baselineskip
    \begin{subfigure}{0.48\textwidth}
        \centering
        \includegraphics[width=\linewidth]{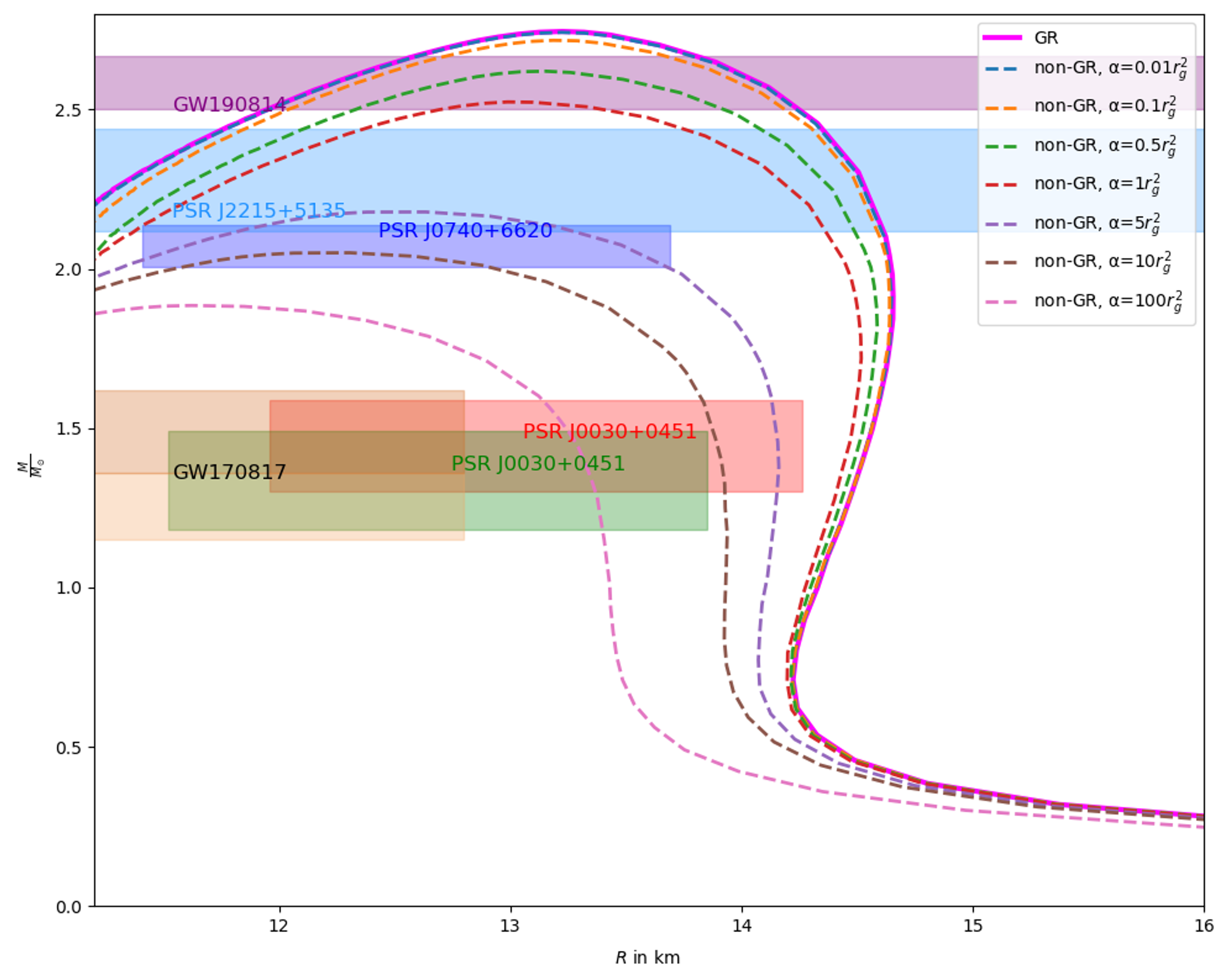}
        \caption{Mass-Radius Diagram for MS1b EoS}
        \label{fig:MRMS1b}
    \end{subfigure}
    \hfill
    \begin{subfigure}{0.48\textwidth}
        \centering
        \includegraphics[width=\linewidth]{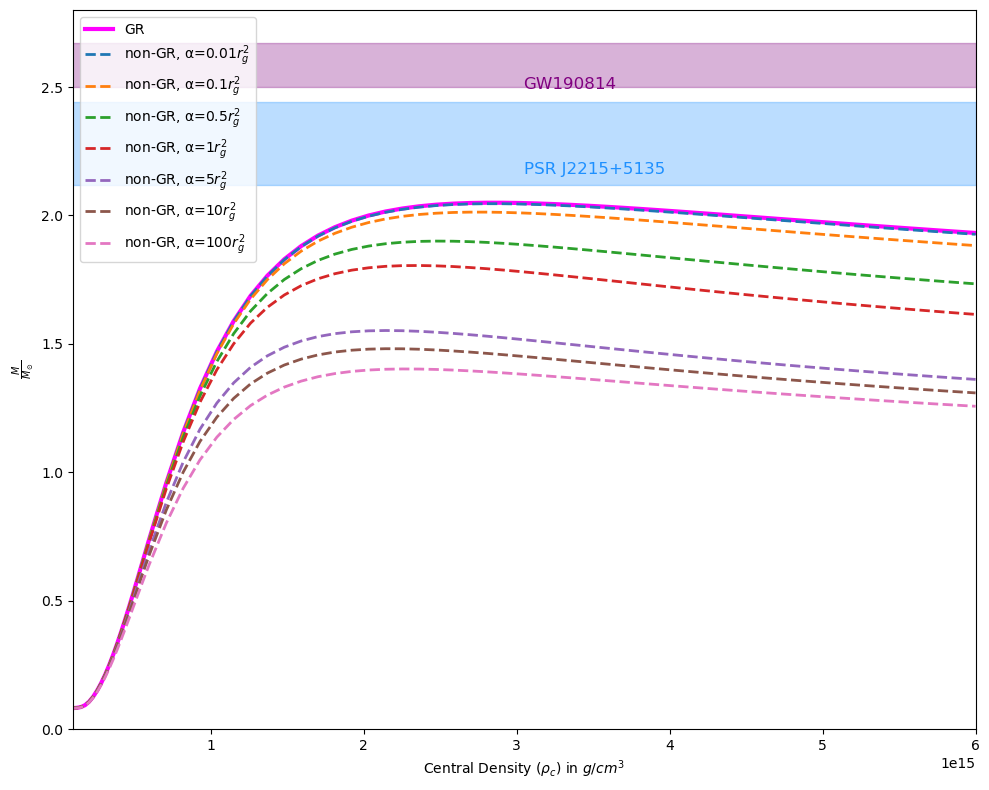}
        \caption{Mass-$\rho_c$ Diagram for SLy EoS}
        \label{fig:MrhocSLy}
    \end{subfigure}
    \caption{Mass-radius and mass-$\rho_c$ diagrams for $f(Q)=Q+\alpha Q^2$ using realistic EoS, SLy, APR4, and MS1b. The values of $\alpha$ range from $0.01r_g^2$ to $100r_g^2$. The plot also includes various observational constraints: GW190814, PSR J2215+5135, PSR J0740+6620, GW170817, and PSR J0030+0451. It can be observed that for small values of $\alpha$, specifically $0.01r_g^2$, the diagram matches the GR case.}
    \label{fig:MRQ2Diagram}
\end{figure}

By repeating the structure calculation of a neutron star over a range of central density values ($\rho_c$), we obtained mass-radius diagrams for several parameter values across all models. The plot also includes various observational constraints from GW events and massive pulsars. GW events detected by LIGO-Virgo, starting with the first event GW170817 \cite{Abbott2017} and subsequent events like GW190814 \cite{Abbott2020}, are believed to involve neutron stars with masses around $2.59 \pm 0.08 M_\odot$. Besides GW events, we also use observational constraints from massive pulsars such as PSR J2215+5135 \cite{Linares2018}, one of the most massive neutron stars, obtained through radio and optical observations with a mass of $2.27^{+0.17}_{-0.15} M_\odot$. Another observational constraint we utilize comes from NICER, such as PSR J0030+0451, with two sets of observational results indicating mass estimates. Miller et al. estimated the mass to be $1.34^{+0.15}_{-0.16} \, M_\odot$ with a radius of $12.71^{+1.14}_{-1.19}$ km \cite{Miller2019}, while Riley et al. estimated the mass to be $1.44^{+0.15}_{-0.14} \, M_\odot$ with a radius of $13.02^{+1.24}_{-1.06}$ km \cite{Riley2019}. The differences between these estimates arise from different assumptions and modeling approaches regarding the thermal emission from the hot spots on the surface of neutron stars. Another massive pulsar from NICER that reported a radius measurement based on fits of rotating hot spot patterns to NICER and X-ray Multi-Mirror (XMM-Newton) X-ray observations is PSR J0740+6620. Miller et al. reported a mass of $2.08 \pm 0.07 \, M_\odot$ and a radius of $13.7^{+2.6}_{-1.5}$ km \cite{Miller2021}, while Riley et al. reported a mass of $2.072^{+0.067}_{-0.066} \, M_\odot$ and a radius of $12.39^{+1.30}_{-0.98}$ km \cite{Riley2021}, using informative priors on pulsar mass, distance, and orbital inclination derived from joint NANOGrav and CHIME/Pulsar wideband radio timing measurements. In this study, we use the observational constraint from \cite{Riley2021}. The results of the mass-radius ($\mathcal{M}-\mathcal{R}$) diagram for the $Q^2$ model show a decrease in mass as the positive $\alpha$ value increases, as illustrated in figure~\ref{fig:MRQ2Diagram}. These results are consistent with previous study \cite{Lin2021}. Furthermore, examining the mass-central density ($\mathcal{M}-\rho_c$) diagram in figure~\ref{fig:MrhocSLy}, the behavior aligns with results from the study using a polytropic EoS, where the mass of star tends to be similar at low $\rho_c$ values. We get also similar result for the other two models. These plots show similarities in the $\mathcal{M}-\mathcal{R}$ and $\mathcal{M}-\rho_c$ diagrams for $f(Q)$ and $f(T)$ gravity theories \cite{Ganiou2017, Iliji2018}. Unfortunately, using the TOV equations (eq.~\ref{eq15}), the solutions obtained are not stable enough to form a star with negative $\alpha$. Therefore, to satisfy the observational constraints in the $Q^2$ model with positive $\alpha$, we need a stiffer EoS like MS1b. When using SLy and APR4 EoS (figure~\ref{fig:MRSLy}, figure~\ref{fig:MRAPR4}), it is observed that only the GW170817 constraint is met. However, when applying a very stiff EoS like MS1b (figure~\ref{fig:MRMS1b}), the mass-radius diagram meets the constraints of GW190814, PSR J2215+5135, PSR J0740+6620, and PSR J0030+0451 across various values of $\alpha$.

\begin{figure}[H]
    \centering
    \begin{subfigure}[b]{0.50\textwidth}
        \centering
        \includegraphics[width=\linewidth]{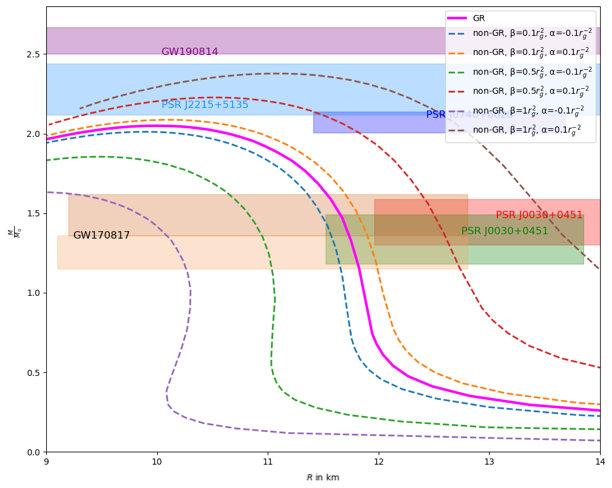}
        \caption{Mass-Radius Diagram for SLy EoS}
        \label{fig:MRSLyE}
    \end{subfigure}%
    \hfill
    \begin{subfigure}[b]{0.48\textwidth}
        \centering
        \includegraphics[width=\linewidth]{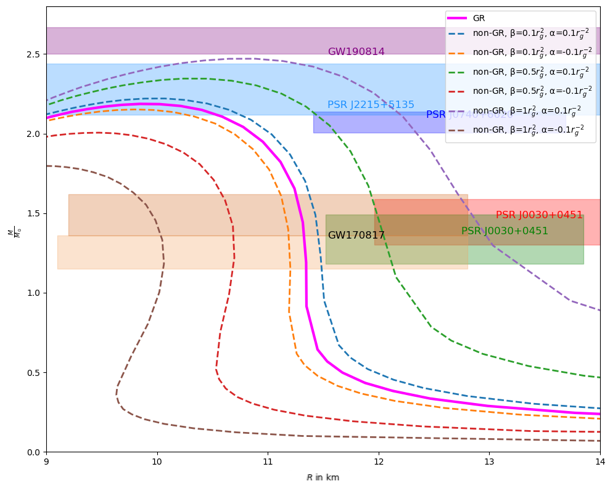}
        \caption{Mass-Radius Diagram for APR4 EoS}
        \label{fig:MRAPRE}
    \end{subfigure}%
    \vskip\baselineskip
    \begin{subfigure}[b]{0.48\textwidth}
        \centering
        \includegraphics[width=\linewidth]{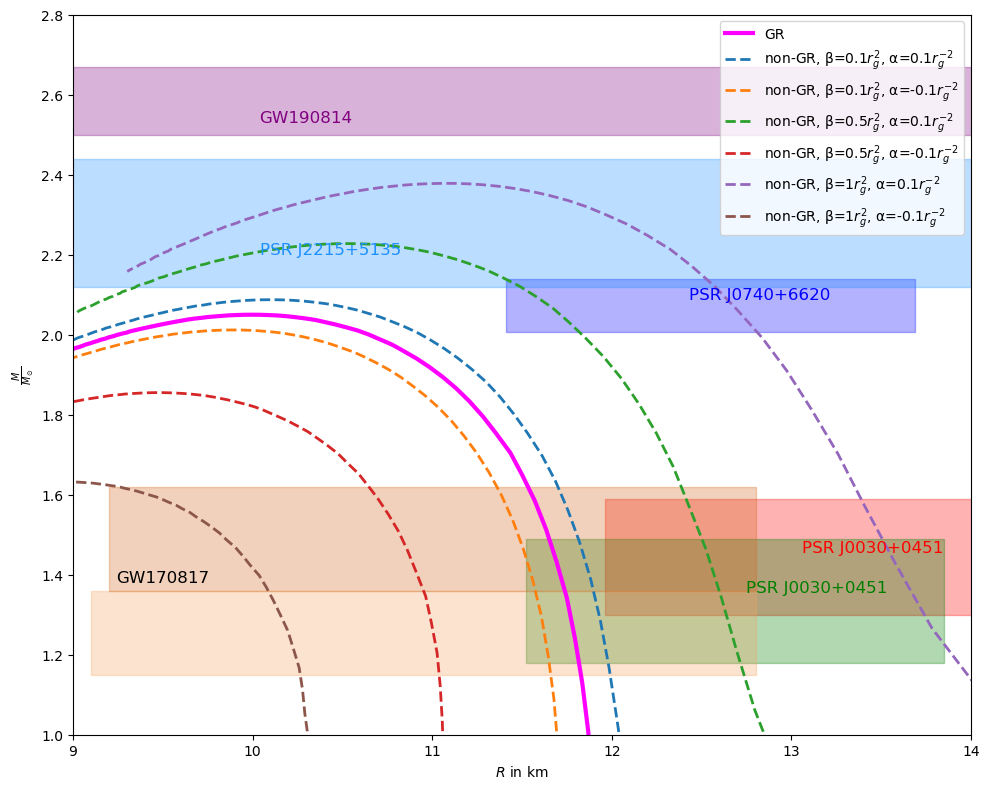}
        \caption{The zoomed-in plot mass-radius diagram for $\alpha= [0.1,-0.1] r_g^{-2} $ and $\beta=[0.1,0.5,0.1]r_g^2$ using SLy EoS}
        \label{fig:MRZoom01}
    \end{subfigure}
    \hfill
    \begin{subfigure}[b]{0.48\textwidth}
        \centering
        \includegraphics[width=\linewidth]{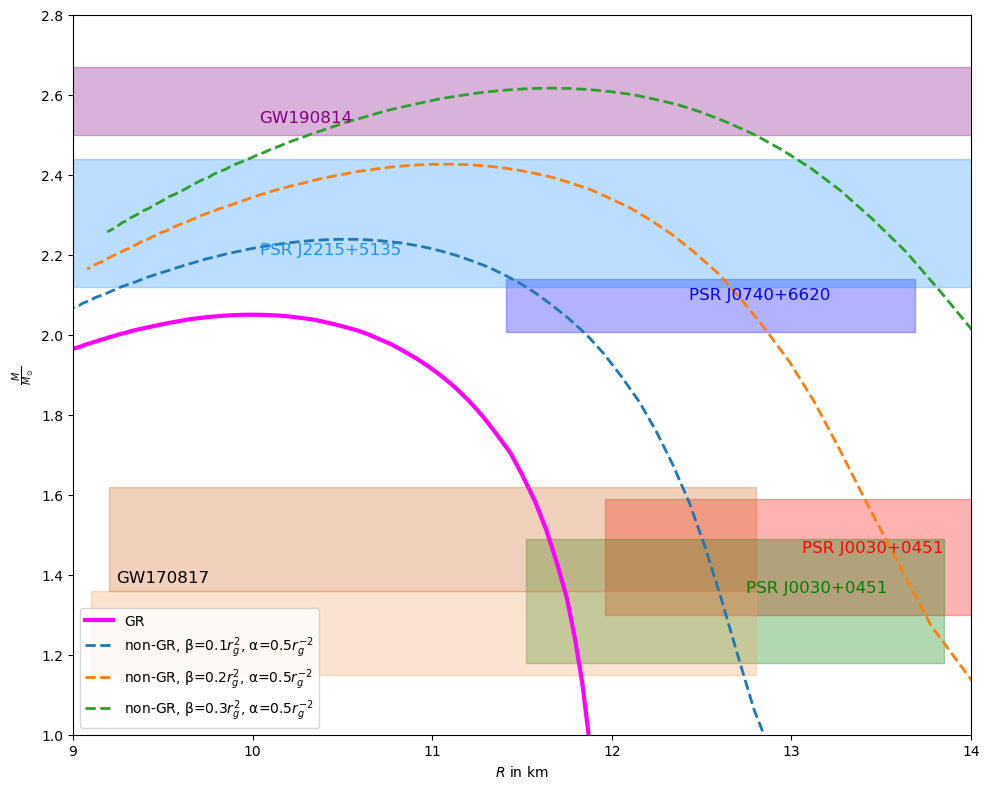}
        \caption{The zoomed-in plot mass-radius diagram for $\alpha= 0.5 r_g^{-2}$ and $\beta=[0.1,0.2,0.3]r_g^2$ using SLy EoS}
        \label{fig:MRZoom05e}
    \end{subfigure}
    
    \caption{Mass-Radius diagrams for $f(Q)=Q+\alpha e^{\beta Q}$ using realistic EoS, SLy and APR4. The plot also includes various observational constraints: GW190814, PSR J2215+5135, PSR J0740+6620, GW170817, and PSR J0030+0451. The zoomed-in plots shows the roles of tuning parameters $\alpha$ and $\beta$. In $\beta$ tuning, it is observed that changes in $\beta$ from 0.1, 0.5, and 1 result in slight variations in the mass of the neutron star, which are not significant. We can compare with $\alpha$ tuning, for $\alpha=0.5r_g^{-2}$, it is possible to achieve a star mass that satisfies the GW190814 constraint. When tuning, it is important to adjust the $\alpha$ and $\beta$ combination so that the star remains stable.}
    \label{fig:MRExpDiagram}
\end{figure}

Because we cannot generate larger stars than GR for negative $\alpha$ in the $Q^2$ model, we try to use another model to compress the correction terms in $f(Q)$, such as $f(Q)=Q+\alpha e^{\beta Q}$ and $f( Q)=Q+\alpha \ln(1-\beta Q)$, where the exponential and logarithmic terms are expected to compress the correction terms and yield stable neutron stars. In figure~\ref{fig:MRExpDiagram}, the $\mathcal{M}-\mathcal{R}$ diagram follows the same pattern as the GR case and shows more stable stars compared to the $Q^2$ plot for higher masses. When $\alpha$ increases positively, a more massive star is obtained. On the other hand, if $\alpha$ decreases negatively, the star becomes less massive. The same behavior is also can be obtained with the $\beta$ parameter if we use same value of $\alpha$: when $\beta$ is positive, a more massive star is obtained, whereas when $\beta$ is negative, we get less massive than in the GR case.

As we said in the previous section, $\alpha$ acts as coarse tuning and $\beta$ as fine tuning. $\alpha$ controls the magnitude of the exponential correction in $f(Q)$, directly affecting the amplitude of the corrections introduced by the exponential term. Therefore, changing the value of $\alpha$ results in significant changes in the structure of the star. The $\beta$ parameter controls the growth rate of the exponential function, influencing the details of the nonlinear corrections in the gravitational field. By adjusting $\alpha$ and $\beta$ together, we can fine-tune the desired structure of the star. 

\begin{figure}[H]
    \centering
    \begin{subfigure}[b]{0.50\textwidth}
        \centering
        \includegraphics[width=\linewidth]{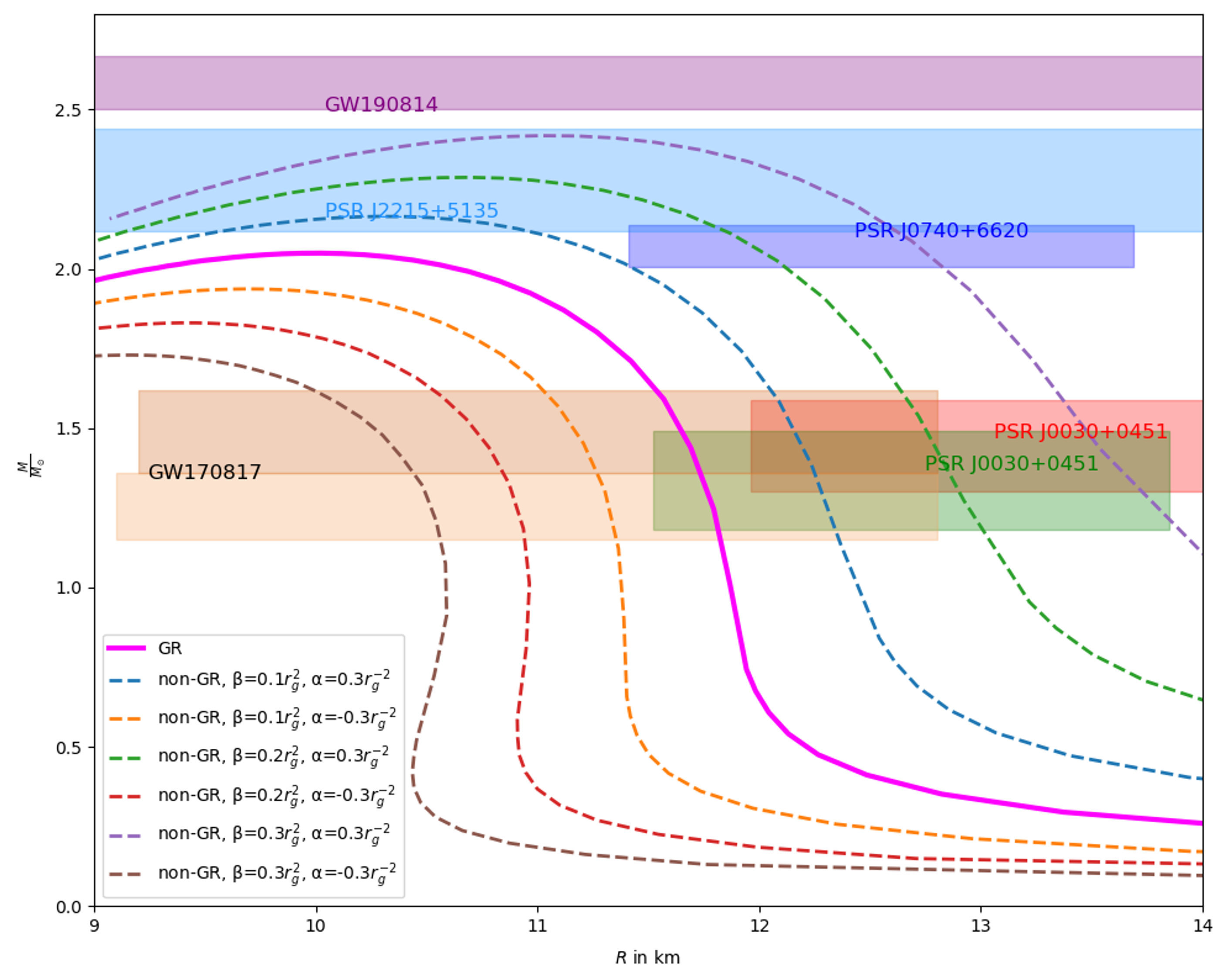}
        \caption{Mass-Radius Diagram for SLy EoS}
        \label{fig:MRSLyL}
    \end{subfigure}%
    \hfill
    \begin{subfigure}[b]{0.48\textwidth}
        \centering
        \includegraphics[width=\linewidth]{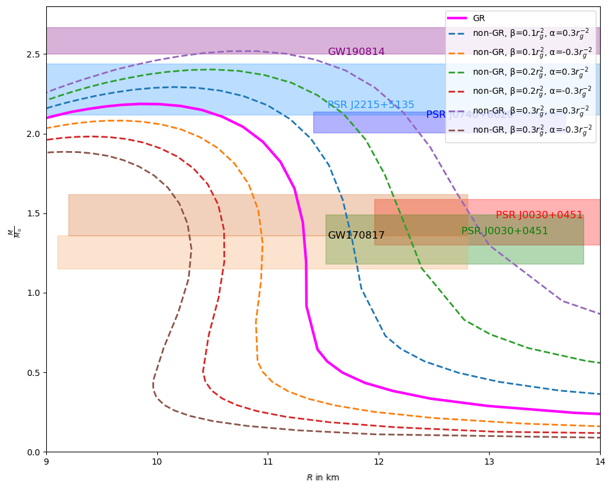}
        \caption{Mass-Radius Diagram for APR4 EoS}
        \label{fig:MRAP4L}
    \end{subfigure}%
    \vskip\baselineskip
    \begin{subfigure}[b]{0.48\textwidth}
        \centering
        \includegraphics[width=\linewidth]{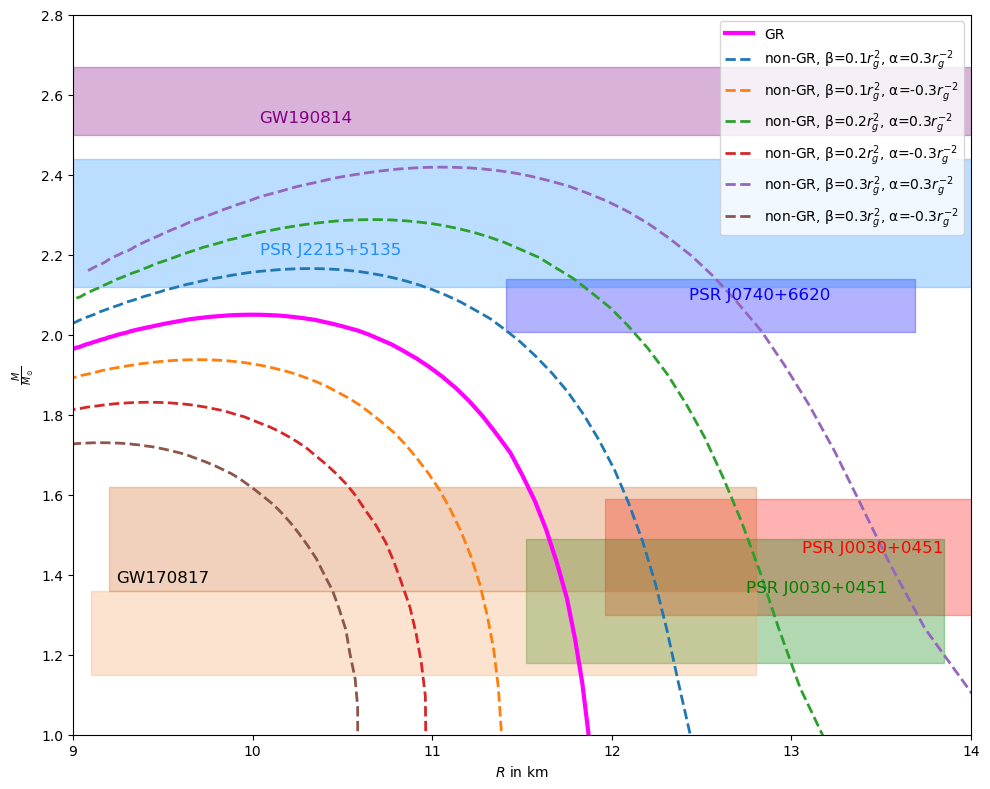}
        \caption{The zoomed-in plot Mass-Radius diagram for $\alpha= [0.3,-0.3]r_g^{-2} $ and $\beta=[0.1,0.2,0.3]r_g^2$ using SLy EoS}
        \label{fig:MRZoom03l}
    \end{subfigure}
    \hfill
    \begin{subfigure}[b]{0.48\textwidth}
        \centering
        \includegraphics[width=\linewidth]{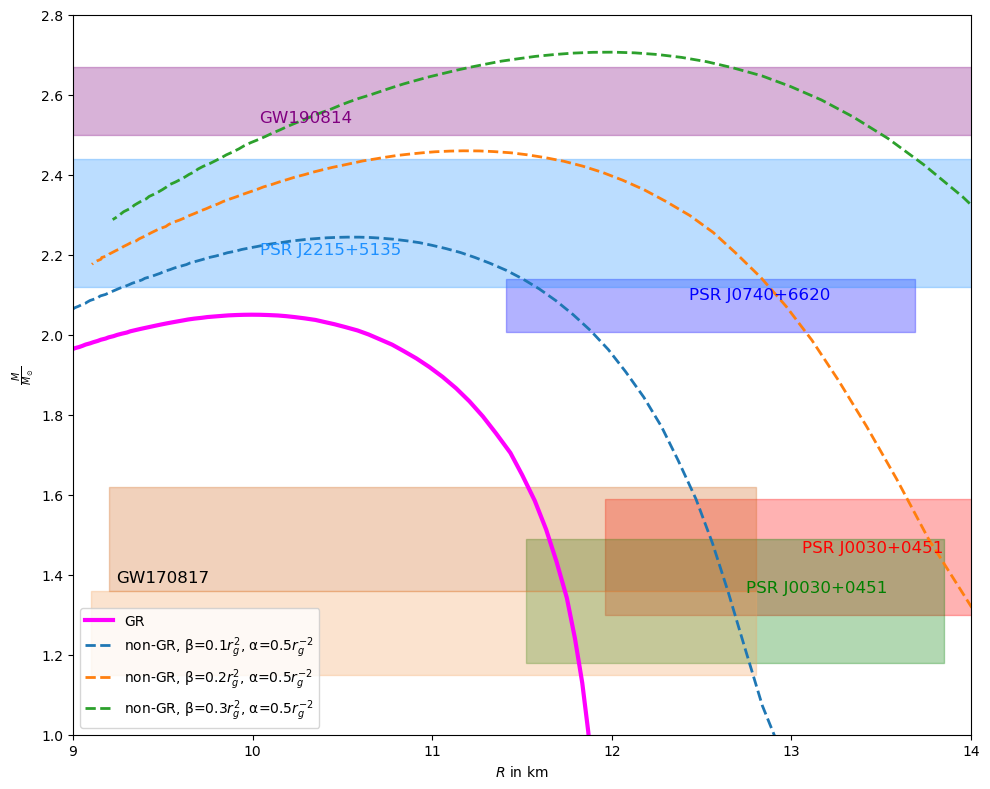}
        \caption{The zoomed-in plot Mass-Radius diagram for $\alpha= 0.5 r_g^{-2}$ and $\beta=[0.1,0.2,0.3]r_g^2$ using SLy EoS}
        \label{fig:MRZoom05l}
    \end{subfigure}
    
    \caption{Mass-Radius diagrams for $f(Q)=Q-\alpha \ln(1-\beta Q)$ using realistic EoS, SLy and APR4. The plot also includes various observational constraints: GW190814, PSR J2215+5135, PSR J0740+6620, GW170817, and PSR J0030+0451. The zoomed-in plots shows the roles of tuning parameters $\alpha$ and $\beta$. In $\beta$ tuning, it is observed that changes in $\beta$ from 0.1 to 1 result in slight variations in the mass of the neutron star, which are not significant. We can compare with $\alpha$ tuning, for $\alpha=0.5r_g^{-2}$, it is possible to achieve a star mass that satisfies the GW190814 constraint.}
    \label{fig:MRLogDiagram}
\end{figure}

\begin{table}[H]
\centering
\setlength{\tabcolsep}{0.5cm}
\renewcommand{\arraystretch}{1.2}
\caption{\label{tab:numresults}
Using the different $\alpha$ and $\beta$ values in the $f(Q)$ model, the maximum mass ($\mathcal{M}_{\text{Max}}$), the radius ($\mathcal{R}$), their ratios represent the compactness ($\mathcal{C}$), and the redshift parameter ($z_s$) of the Neutron stars for the SLy EoS.}
\begin{tabular}{|c|c c c c c c|}  
\hline
$\bm{f\left(Q\right)}$ \textbf{Model} & $\bm{\alpha}$ & $\bm{\beta}$ & $\bm{\mathcal{M}_{Max}}$ & $\bm{\mathcal{R}}$ & $\bm{\mathcal{C}}$ & $\bm{z_s}$ \\[0.3cm]
\hline\hline
\multirow{8}{*}{\shortstack{$\bm{Q+\alpha Q^2}$}} 
& \multicolumn{1}{r}{\textcolor{red}{GR}}    & \multicolumn{1}{c}{}    & \textcolor{red}{2.05} & \textcolor{red}{9.995} & \textcolor{red}{0.205} & \textcolor{red}{0.302} \\
\cline{2-7}
& \multicolumn{1}{r}{0.01}  & \multirow{7}{*}{}      & 2.046 & 9.989 & 0.205 & 0.302 \\
& \multicolumn{1}{r}{0.1}   &                         & 2.013 & 10.044 & 0.200 & 0.291 \\
& \multicolumn{1}{r}{0.5}   &                         & 1.900 & 10.032 & 0.189 & 0.268 \\
& \multicolumn{1}{r}{1}     &                         & 1.805 & 10.029 & 0.180 & 0.250 \\
& \multicolumn{1}{r}{5}     &                         & 1.552 & 9.545 & 0.163 & 0.218 \\
& \multicolumn{1}{r}{10}    &                         & 1.481 & 9.346 & 0.158 & 0.209 \\
& \multicolumn{1}{r}{100}   &                         & 1.402 & 8.987 & 0.156 & 0.206 \\
\hline
\multirow{12}{*}{\shortstack{$\bm{Q+\alpha e^{\beta Q}}$}} 
& \multirow{3}{*}{-0.1} & 0.1 & 2.012 & 9.8895  & 0.203 & 0.297 \\
&                      & 0.5 & 1.855 & 9.460 & 0.196 & 0.282 \\
&                      & 1   & 1.632 & 8.947 & 0.182 & 0.254 \\
\cline{2-7}
& \multicolumn{1}{r}{\textcolor{red}{GR}}    & \multicolumn{1}{c}{}    & \textcolor{red}{2.05} & \textcolor{red}{9.995} & \textcolor{red}{0.205} & \textcolor{red}{0.302} \\
\cline{2-7}
& \multirow{3}{*}{0.1} & 0.1 & 2.087 & 10.098 & 0.207 & 0.306 \\
&                      & 0.5 & 2.227 & 10.531 & 0.211 & 0.315 \\
&                      & 1   & 2.378 & 11.053 & 0.215 & 0.325 \\
\cline{2-7}
& \multirow{3}{*}{0.5} & 0.1 & 2.238 & 10.515   & 0.213 & 0.320 \\
&                      & 0.2 & 2.426 & 11.048 & 0.220 & 0.336 \\
&                      & 0.3 & 2.616 & 11.663 & 0.224 & 0.346 \\
\cline{2-7}
& \multirow{2}{*}{1}   & 0.1 & 2.429 & 11.078 & 0.219 & 0.334 \\
&                      & 0.2 & 2.826 & 12.294 & 0.230 & 0.361 \\
\hline
\multirow{11}{*}{\shortstack{$\bm{Q-\alpha\ln\left(1-\beta Q\right)}$}}
& \multirow{3}{*}{-0.3} & 0.1 & 1.937 & 9.702  & 0.200   & 0.291 \\
&                      & 0.2 & 1.831 & 9.406 & 0.195 & 0.280 \\
&                      & 0.3 & 1.73  & 9.164 & 0.189 & 0.268 \\
\cline{2-7}
& \multicolumn{1}{r}{\textcolor{red}{GR}}    & \multicolumn{1}{c}{}    & \textcolor{red}{2.05} & \textcolor{red}{0.995} & \textcolor{red}{0.205} & \textcolor{red}{0.302} \\
\cline{2-7}
& \multirow{3}{*}{0.3} & 0.1 & 2.165 & 10.310  & 0.210  & 0.313 \\
&                      & 0.2 & 2.288 & 10.690  & 0.214 & 0.322 \\
&                      & 0.3 & 2.418 & 11.027 & 0.219 & 0.334 \\
\cline{2-7}
& \multirow{3}{*}{0.5} & 0.1 & 2.244 & 10.529 & 0.213 & 0.320 \\
&                      & 0.2 & 2.459 & 11.172 & 0.220  & 0.336 \\
&                      & 0.3 & 2.706 & 12.009 & 0.225 & 0.348 \\
\cline{2-7}
& \multicolumn{1}{r}{1} & 0.1 & 2.447 & 11.206 & 0.218 & 0.332 \\
&                      & 0.2 & 2.968  & 12.964 & 0.229 & 0.358 \\
\hline
\end{tabular}

\end{table}

\begin{figure}[H]
    \centering
    \begin{subfigure}[b]{0.32\textwidth}
        \centering
        \includegraphics[width=\linewidth]{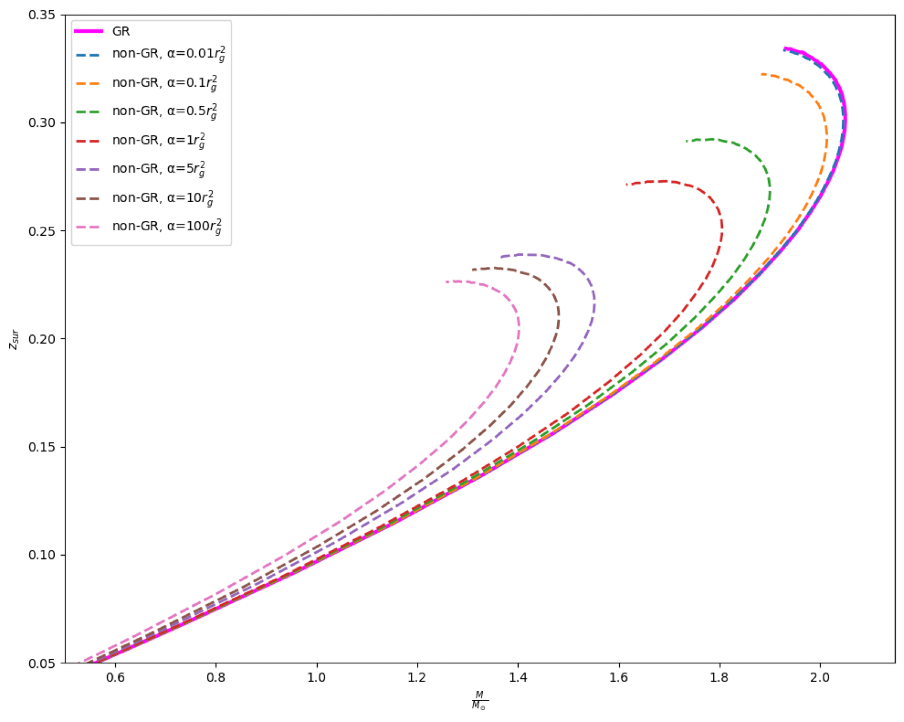}
        \caption{$f(Q)=Q+\alpha Q^2$}
    \end{subfigure}%
    \hfill
    \begin{subfigure}[b]{0.32\textwidth}
        \centering
        \includegraphics[width=\linewidth]{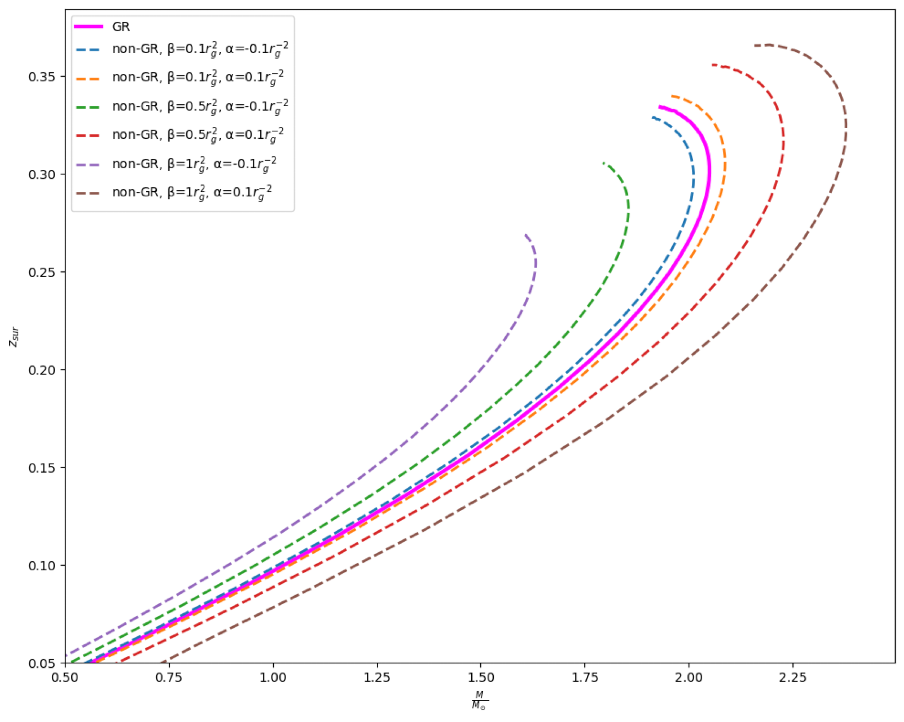}
        \caption{$f(Q)=Q+\alpha e^{\beta Q}$}
    \end{subfigure}%
    \hfill
    \begin{subfigure}[b]{0.32\textwidth}
        \centering
        \includegraphics[width=\linewidth]{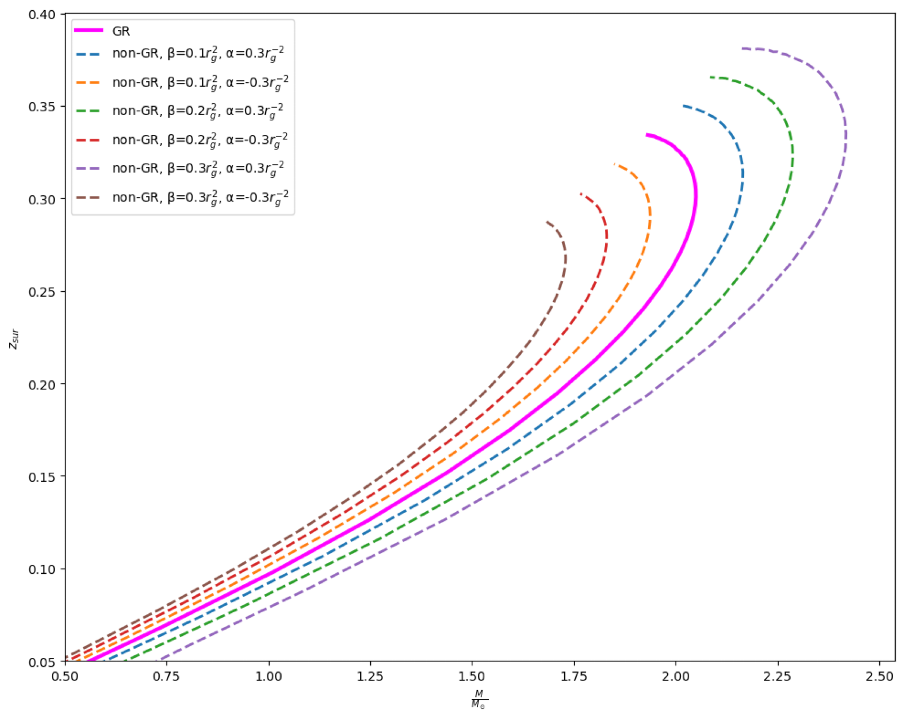}
        \caption{$f(Q)=Q-\alpha \log (1-\beta Q)$}
    \end{subfigure}
    \caption{The surface gravitational redshift $z_s$ function of $M/M_\odot$ for each $f(Q)$ model. In this plot, we use SLy EoS and some parameter $\alpha$ and $\beta$. From this plot, we can see $z_s$ fulfill constraint from \cite{Buch1959}.}
    \label{fig:zs}
\end{figure}

As shown in figures~\ref{fig:MRZoom01}, when adjusting the $\beta$ parameter with $\alpha=0.1r_g^{-2}$, there is a smooth change in the mass of star. In contrast, in figure~\ref{fig:MRZoom05e} with $\alpha=0.5r_g^{-2}$ at the same $\beta$ value, the change in the mass of star is more significant and can reach the observational constraint for the possibily most massive star GW190814. The significant role of $\alpha$ is also visible when returning the $f(Q)$ equation to GR at very small alpha values such as $\alpha=0.01 r_g^{-2}$ as in figure~\ref{fig:ASLyQexp},~\ref{fig:BSLyQexp},~\ref{fig:pSLyQexp} and ~\ref{fig:QSLyQexp}.

In the model $f(Q)=Q-\alpha \ln(1-\beta Q)$, we obtained results that has similar behavior to the exponential model. As shown in figure~\ref{fig:MRLogDiagram}, with positive $\alpha$, we obtain more massive stars, and with negative $\alpha$, we obtain less massive stars than in the GR case. At certain parameter values, it can be seen that the diagram meets GW190814, PSR J2215+5135, PSR J0740+6620, GW170817, and PSR J0030+0451 constraints.

Both $\alpha$ and $\beta$ play important roles in this model as we explained in the previous section. $\alpha$ controls the effect of the logarithmic term, while $\beta$ determines the critical point and the rate of decrease of the logarithmic function. It is evident that when $\beta$ is very small, the solution converges to the GR solution. The effect of these two parameters can also be seen in figure~\ref{fig:MRZoom03l} and figure~\ref{fig:MRZoom05l}. When adjusting $\alpha=0.5r_g^{-2}$ with $\beta=0.3 r_g^2$, we can achieve sufficiently massive stars that meet the GW190814 constraint. In both models, using $\alpha=2r_g^{-2}$, we can get massive stars $>$2.75$M_\odot$ at small $\beta$ values such as $\beta=0.1 r_g^2$. However, we encounter numerical difficulties at low central densities for this parameter value. Careful tuning of these two parameters is required for both models because the star can become very unstable if the exponential or logarithmic terms are too large. 

The other neutron star properties that we can obtain from mass-radius relation are compactness ($\mathcal{C}$) and surface gravitational redshift ($z_s$). Now, $\mathcal{C}$ can be defined as,
\begin{equation}\label{compactness}
\mathcal{C}=\mathcal{M}/\mathcal{R}.
\end{equation}
Buchdal \cite{Buch1959} has provided the upper limit for the compactness of neutron star, which will remain stable if the value $\leq \frac{4}{9}$. For $z_s$, defined as $\frac{1}{\sqrt{-g_{tt}}}-1$, the $\Lambda$ term in $g_{tt}$ at the surface of the star in the SdS solution can be ignored, as discussed in subsection \ref{subsec:4.2}. The $\Lambda$ term becomes zero for the quadratic and logarithmic models, and can be neglected for the exponential model due to the $r_g$ constant suppresses the $\Lambda$ term. Therefore, we can use the same definition of $z_s$ from GR as,
\begin{equation}
    z_s=\frac{1-\sqrt{1-2\mathcal{C}}}{\sqrt{1-2\mathcal{C}}}.
\end{equation}
According to Refs. \cite{Buch1959} and \cite{N.Straumann}, the surface gravitational redshift must satisfy $z_s \leq 2$ for neutron stars. Both $\mathcal{C}$ and $z_s$ have been shown in Table~\ref{tab:numresults} and figure~\ref{fig:zs}. As shown in Table~\ref{tab:numresults}, the two parameters, $\alpha$ and $\beta$, have different roles in star compactness in the exponential and logarithmic models. In both models, $\alpha$ and $\beta$ affect compactness, with $\alpha$ acting as coarse tuning having a larger effect than $\beta$, which acts as fine tuning. Compactness increases with the increasing values of both parameters, and vice versa. The $z_s$ for both the models are also consistent with the $\mathcal{M}-\mathcal{R}$ diagram, where positive $\alpha$ results in a larger $z_s$ and negative $\alpha$ results in a smaller $z_s$. From figure~\ref{fig:zs}, we can see that at low mass, $z_s$ values do not differ significantly, but at higher mass, the deviation in $z_s$ due to parameter differences becomes more pronounced. For the quadratic model, we can also observe how compactness and $z_s$ decrease with increasing $\alpha$. From the table, we can conclude that all configurations satisfy the neutron star compactness and $z_s$ limits.


\section{Discussion}\label{Sec:V}
After obtaining the solutions for the structure of each star and the mass-radius relationship of neutron stars in covariant $f(Q)$ gravity, we can observe that nonmetricity has an important role in star formation. Looking at the results from $f(Q)=Q+\alpha Q^2$, we find that the star becomes lighter as $\alpha$ increases. This reduction in mass is accompanied by a decrease in the profile of $Q(r)$, indicating a correlation between $Q$ and the ability of star to acquire matter. Unfortunately, using the TOV equations [eq.~\eqref{eq15}], the solutions obtained are not stable enough to form a star with negative $\alpha$. Furthermore, we have also tried using higher-order corrections, such as $f(Q)=Q + \alpha Q^2 + \beta Q^n$, but the results remain the same. These models still cannot generate higher masses, indicating that they are unable to compress the $Q^2$ corrections sufficiently to achieve higher mass configurations. This limitation may be due to the inability of the higher-order terms to effectively counteract the dominant $Q^2$ term, leading to insufficient structural changes to support a more massive star. Previous works on $f(Q)$ and $f(T)$ gravity \cite{Iliji2018, Lin2021} have shown that for negative $\alpha$, a critical point for stability cannot be found. For positive $\alpha$, the modified gravity terms tend to stabilize the star by allowing configurations with less massive stars compared to the GR case. When $\alpha$ is negative, the nonmetricity function $f(Q)$ introduces significant deviations from GR, affecting the equilibrium of the star. These changes can indirectly affect the structure of star, allowing it to hold more mass. However, this alteration in structure can make the star unstable, especially when the central density and energy density exceed a certain threshold. This instability prevents the formation of stable neutron stars with negative $\alpha$, as the changes to the structure of star disrupt the balance needed to keep it stable. We found different results with the other two models. In these models, the additional terms in $f(Q)$ can effectively compress the value of $|Q|$, allowing the both more massive and less massive stars without issues. However, the parameters $\alpha$ and $\beta$ are important for maintaining the stability of the structure of star.

Moreover, if we look at the $\mathcal{M}-\rho_c$ diagram in figure \ref{fig:MrhocSLy}, the modification effect of gravity is not significant at low central density, so the results are similar to GR. As the central density increases, modified $f(Q)$ begin to affect the structure of star, causing changes in mass and radius. This corresponds to \cite{Barker2024}, which indicates that at low energies, the effects of modified gravity will not be very visible because the solution obtained will return to GR.

So, from the solutions of all models, we can opine that the nonmetricity in the star affects the geometry of the interior structure of the star, which consists of perfect fluid. Thus affecting the distribution of pressure and matter. When the profile $Q(r)$ starts to deviate due to changes in the parameters $\alpha$ and $\beta$, the profile $B(r)$ also starts to deviate. The metric tensor coefficient $e^{B(r)}$ is related to the mass function of the star through the equation $e^{-B(r)} = 1 - \frac{2m(r)}{r}$. Therefore, any deviations in $B(r)$ will directly affect the interior mass function $m(r)$. When $B(r)$ deviates due to $Q(r)$, the star can either accommodate more matter or less matter, impacting the total mass. This change in matter distribution is also accompanied by pressure deviations due to changes in $A(r)$. Strong gravity in the core pulls the matter of star inward, attempting to compress it, while internal pressure generated by nuclear reactions and degeneracy pressure counteracts this gravitational pull to prevent collapse. The deviation $A(r)$ due to $Q(r)$ changes this balance by changing the internal pressure. If the pressure increases, the star can support more mass and remain stable. Conversely, if the pressure decreases, the star loses matter, leading to a decrease in total mass. Thus, the stability of the star is closely tied to the deviations in both $A(r)$ and $B(r)$ caused by $Q(r)$.

We can find a similar phenomenon in how $f(Q)$ affects the pressure distribution in strange stars \cite{Lohakare2023}, where the radial and tangential pressures inside the star are influenced by nonmetricity. In these stars, anisotropic fluids are used, and as $|Q|$ increases, the nonmetricity scalar changes the interior structure of the star, leading to an increase in both radial and tangential pressures. A similar effect can be found in other references, such as the $f(T)$ model \cite{Lin2021a} or previous studies on $f(Q)$ \cite{Lin2021} that used a polytropic EoS for their calculations. These studies explain how torsion in $f(T)$ or nonmetricity in $f(Q)$ affects the geometric fluid, impacting the pressure and enabling the star to accommodate more matter. Therefore, this scenario is also possible in the context of our study.


\section{Conclusions}\label{Sec:VI}
In this paper, we have studied neutron stars in covariant $f(Q)$ gravity using three modified $f(Q)$ models: $f(Q)=Q+\alpha Q^2$; $f(Q)=Q+\alpha e^{\beta Q}$; and $f(Q)=Q-\alpha \ln(1-\beta Q)$. By using piecewise polytrope EoS, we obtained the metric profiles $A(r)$ and $B(r)$, nonmetricity $Q(r)$, and pressure $p(r)$ in the interior of the star, which match the exterior SdS solution outside the star. We also calculated $\mathcal{M}-\mathcal{R}$ diagram of neutron stars. The parameters $\alpha$ and $\beta$ have their respective roles in each model in influencing the structure of the star.

In the model $f(Q)=Q+\alpha Q^2$, the profile $Q(r)$ decreases as the positive value of $\alpha$ increases. This affects the profiles of $A(r)$ and $B(r)$, influencing the mass and compactness of the star. Our results show that as $\alpha$ increases, the star becomes less massive and less compact. Previous studies also have shown that with negative $\alpha$, it is possible to obtain more massive neutron stars. However, with the TOV equations we used, we could not generate stable neutron stars with negative $\alpha$, which we suspect is due to the instability in the $Q'$ term when controlling the $\alpha Q^2$ correction. In the models $f(Q)=Q+\alpha e^{\beta Q}$ and $f(Q)=Q-\alpha \ln(1-\beta Q)$, we successfully obtained neutron stars that are more massive and compact compared to those in GR. The parameter $\alpha$ controls the influence of the correction term, while the parameter $\beta$ acts as a fine-tuning parameter for the exponential or logarithmic growth rate. By carefully tuning these parameters to small values, we can modify the structure of star to achieve more massive or less massive stars. Using SLy and APR4 EoS, the resulting $\mathcal{M}-\mathcal{R}$ diagrams meet the constraints from GW190814, PSR J2215+5135, PSR J0740+6620, GW170817, and PSR J0030+0451. For example, in the logarithmic model, with $\beta=0.3r_g^2$ and $\alpha=0.3r_g^{-2}$, the $\mathcal{M}-\mathcal{R}$ diagram meets all observational constraints when using the APR4 EoS. In contrast to the $f(Q)=Q +\alpha Q^2$ model with positive $\alpha$ which requires a very stiff EoS such as MS1b to satisfy observational constraints. Careful tuning of $\alpha$ and $\beta$ parameters is necessary to avoid stellar instability because achieving higher masses just requires small parameter values. Additionally, as shown in Table \ref{tab:numresults}, all configurations satisfy the neutron star compactness and $z_s$ limit, $\mathcal{C}\leq \frac{4}{9}$ and $z_s \leq 2$.

From all results, it can be observed that $Q$ has important role in the structure of neutron star matter. In summary, the nonmetricity affects the internal geometry of the star, which in turn affects the density, pressure, and overall stability of the neutron star. This enables the star to accommodate more matter and withstand a heavier mass. It will also be interesting to consider hair solution for solving neutron stars in $f(Q)$ gravity. This approach may provide more comprehensive results and improve our understanding of the modifications introduced by gravity $f(Q)$ in the context of neutron star structure. This scenario has also been studied in various works related to $f(R)$ gravity in neutron stars \cite{Astashenok2019, Feola2020, Numajiri2023}. Hair solution in $f(R)$ gravity has been shown to introduce additional stability and modify the exterior and interior solutions of stars, leading to a more comprehensive understanding of stellar structures under modified gravity theories. Moreover, considering more general solutions, as demonstrated in studies on black holes in $f(Q)$ gravity \cite{D'Ambrosio2022}, by constructing the most general static and spherically symmetric forms of the metric and the affine connection, could provide further insights into the structure and stability of neutron stars.


\acknowledgments
The authors thank N. Yoshioka for the useful discussion. MAA would also like to thank M. D. Danarianto for his helpful discussion regarding numerical methods on neutron stars. BM thanks IUCAA, Pune (India) for providing support in the form of an academic visit during which this work is accomplished. SAN acknowledges the financial support provided by Hiroshima University, Japan through Japan Student Services Organization (JASSO) Fellowship to carry out the research work.


\appendix
\section{Energy-Momentum Conservation}\label{AppendixA}
One of the issues in $f(Q)$ gravity within the spherically symmetry metric is the conservation of energy-momentum. In accordance with Refs. \cite{Zhao2022} and \cite{De2023}, a constraint arises on the left-hand side when the covariant derivative is applied to eq.~(\ref{eq7}). Let's redefine eq.~(\ref{eq7}) where the left side represents the gravitational part as $E_{\mu\nu}$, and the right side represents the matter part, $\mathcal{T}_{\mu\nu}$. 
\begin{align}
    E_{\mu\nu} \equiv f_{Q}\mathring{G}_{\mu\nu}+\frac{1}{2}g_{\mu\nu}(Qf_{Q}-f)+2f_{QQ}P^{\lambda}_{~~\mu\nu}\mathring{\nabla}_{\lambda}Q &= \kappa \mathcal{T}_{\mu \nu} .
\end{align}
As discussed in \cite{Zhao2022}, the right side, under the assumption of energy-momentum conservation, can easily become zero when the covariant derivative is applied, $\mathring{\nabla}_\mu \mathcal{T}^{\mu \nu} = 0$. However, the covariant derivative of the left side, representing the gravitational part, will give a constraint as:
\begin{align}
    \frac{(e^{B}-1)(4+rA'+rB')+2rB'}{2r^2}f_Q' + \frac{(e^{B}-1)}{r}f_Q''&=0 ,\label{apA}
\end{align}
where $ f_Q' = f_{QQ} \frac{dQ}{dr} = f_{QQ} Q'$ and $ f_Q'' = f_{QQQ} \left( \frac{dQ}{dr} \right)^2 + f_{QQ} \frac{d^2Q}{dr^2}=f_{QQQ} Q'^2 + f_{QQ} Q''$. So eq.~(\ref{apA}) becomes
\begin{align}
    \left[\frac{(e^{B}-1)(4+rA'+rB')+2rB'}{2r^2}\right]f_{QQ} Q'+\frac{(e^{B}-1)}{r}\left(f_{QQQ} Q'^2 + f_{QQ} Q''\right)&=0\nonumber\\
    f_{QQ}\left[\frac{\left( e^{B} - 1 \right) rA' - \left( e^{B} + 1 \right) rB' + 4 \left( e^{B} - 1 \right)}{2 r^2} Q' + \frac{e^{B} - 1}{r} Q'' \right]&\nonumber\\
    + f_{QQQ}\left[\frac{e^{B} - 1}{r}Q'^2\right]&=0\nonumber\\
    f_{QQ}\Phi_r +f_{QQQ}\Psi_r&=0\label{emu}
\end{align}

\begin{figure}[H]
    \centering
    \begin{subfigure}[b]{0.32\textwidth}
        \centering
        \includegraphics[width=\linewidth]{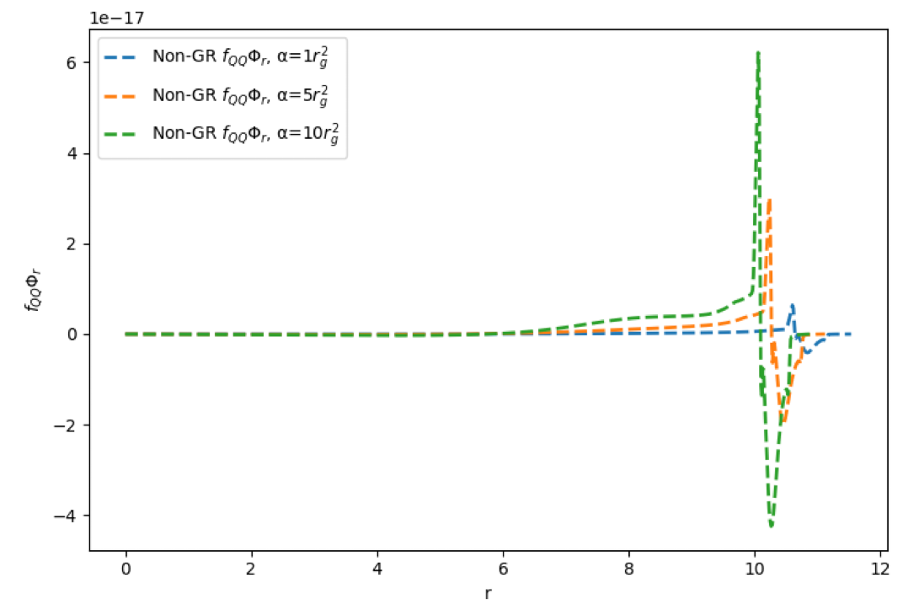}
        \caption{$f_{QQ}\Phi_r$ of $Q^2$ model}
    \end{subfigure}%
    \hfill
    \begin{subfigure}[b]{0.32\textwidth}
        \centering
        \includegraphics[width=\linewidth]{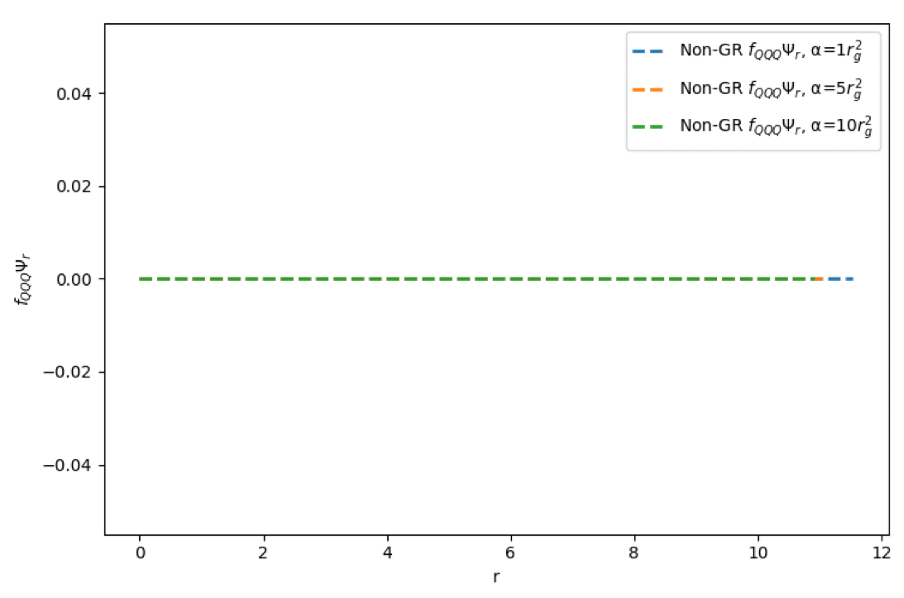}
        \caption{$f_{QQQ}\Psi_r$ of $Q^2$ model}
    \end{subfigure}%
    \hfill
    \begin{subfigure}[b]{0.32\textwidth}
        \centering
        \includegraphics[width=\linewidth]{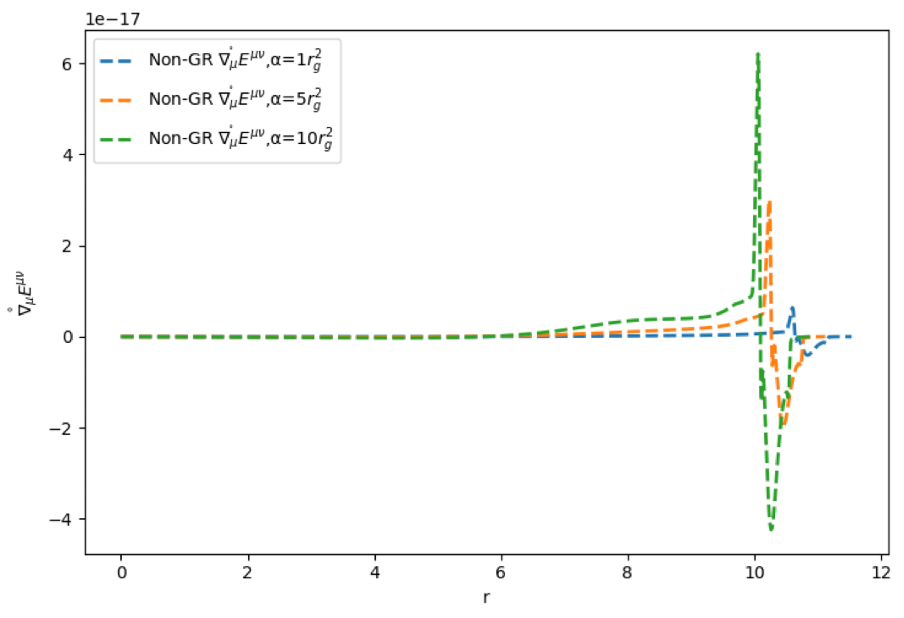}
        \caption{$\mathring{\nabla}_\mu E^{\mu\nu}$ of $\alpha Q^2$ model}
    \end{subfigure}
    \hfill
    \begin{subfigure}[b]{0.32\textwidth}
        \centering
        \includegraphics[width=\linewidth]{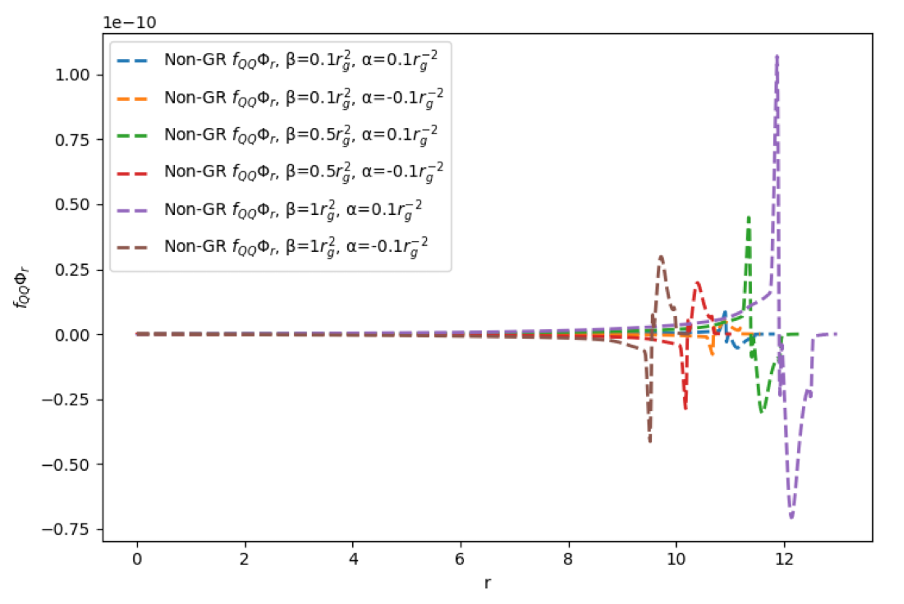}
        \caption{$f_{QQ}\Phi_r$ of $\alpha e^{\beta Q}$ model}
    \end{subfigure}%
    \hfill
    \begin{subfigure}[b]{0.32\textwidth}
        \centering
        \includegraphics[width=\linewidth]{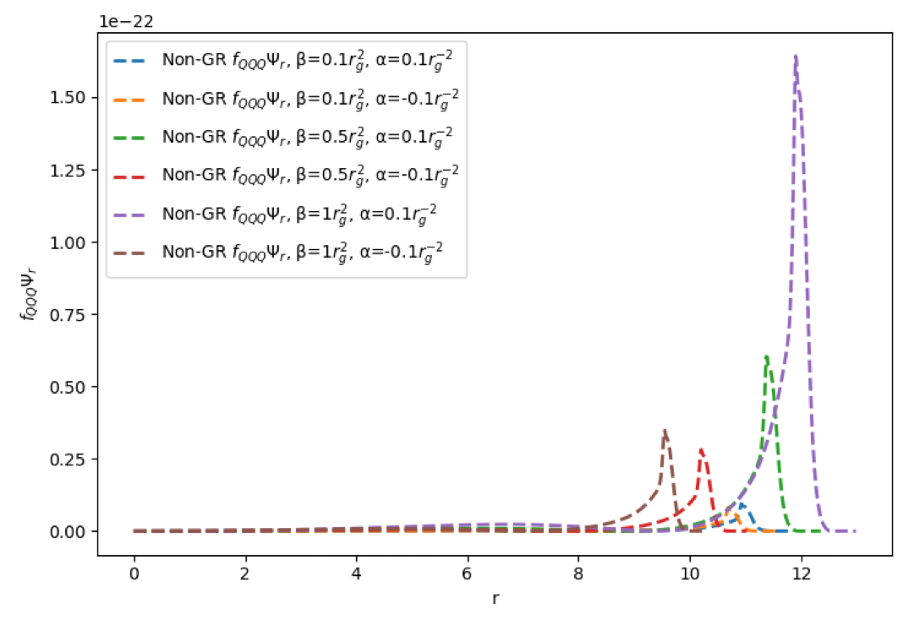}
        \caption{$f_{QQQ}\Psi_r$ of $\alpha e^{\beta Q}$ model}
    \end{subfigure}%
    \hfill
    \begin{subfigure}[b]{0.32\textwidth}
        \centering
        \includegraphics[width=\linewidth]{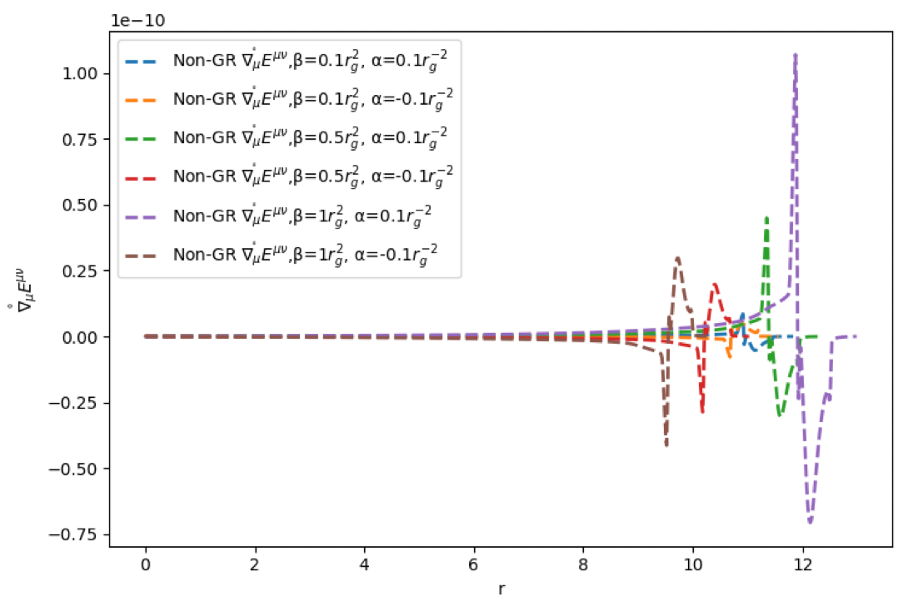}
        \caption{$\mathring{\nabla}_\mu E^{\mu\nu}$ of $\alpha e^{\beta Q}$ model}
    \end{subfigure}
    \hfill
    \begin{subfigure}[b]{0.32\textwidth}
        \centering
        \includegraphics[width=\linewidth]{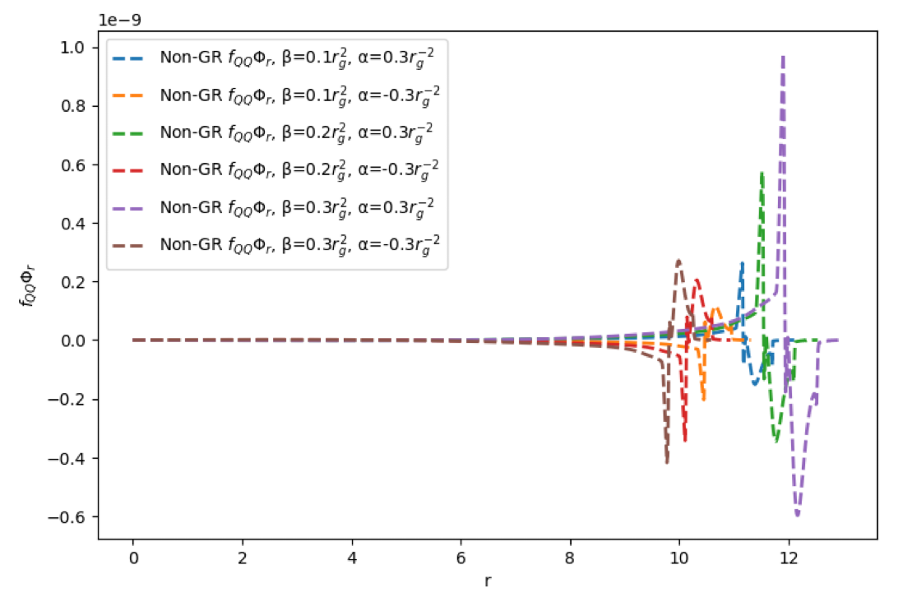}
        \caption{$f_{QQ}\Phi_r$ of $\alpha \log (1-\beta Q)$ model}
    \end{subfigure}%
    \hfill
    \begin{subfigure}[b]{0.32\textwidth}
        \centering
        \includegraphics[width=\linewidth]{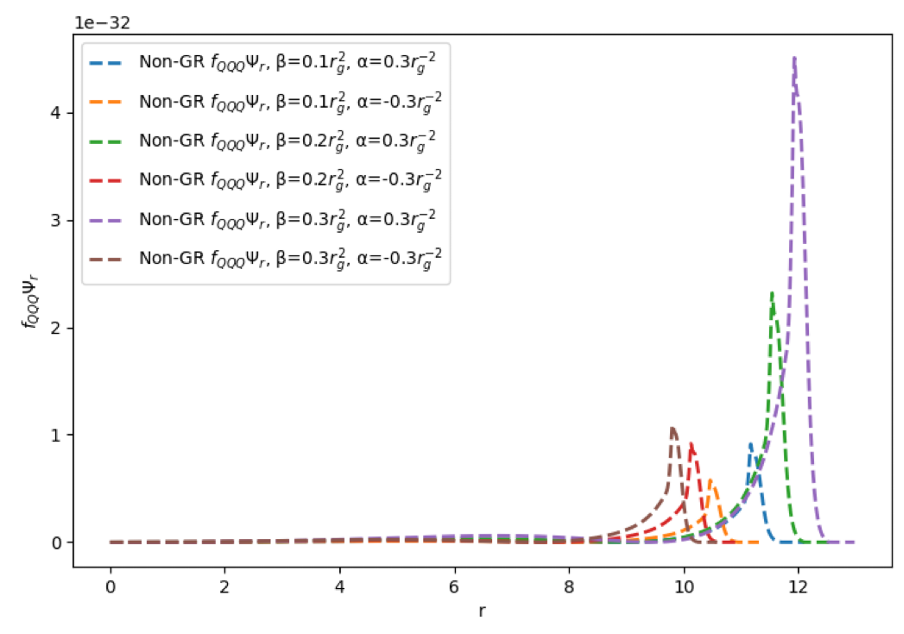}
        \caption{$f_{QQQ}\Psi_r$ of $\alpha \log (1-\beta Q)$ model}
    \end{subfigure}%
    \hfill
    \begin{subfigure}[b]{0.32\textwidth}
        \centering
        \includegraphics[width=\linewidth]{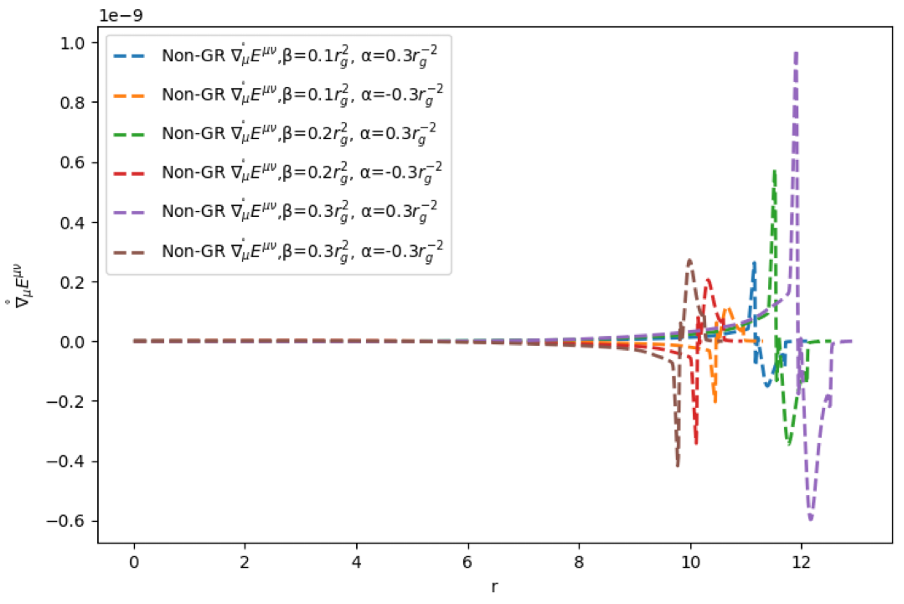}
        \caption{$\mathring{\nabla}_\mu E^{\mu\nu}$ of $\alpha \log (1-\beta Q)$ model}
    \end{subfigure}
    \caption{$f_{QQ}\Phi_r$, $f_{QQQ}\Psi_r$, and $\mathring{\nabla}_\mu E^{\mu\nu}$ for each model using SLy EoS. In this plot, we use $\rho_c = 1 \times 10^{15}$ g/cm$^3$. Note that the quadratic model has no $f_{QQQ}$ term. These plots demonstrate that $\mathring{\nabla}_\mu E^{\mu\nu}$ for each model is very small, with the largest value in the logarithmic model reaching the order of $10^{-9}$. Additionally, these results are consistent with analytical calculations, showing that at the core and surface of the star, $\mathring{\nabla}_\mu E^{\mu\nu} = 0$ due to the absence of nonmetricity.}
    \label{fig:emu}
\end{figure}

We can see that $\mathring{\nabla}_\mu E^{\mu\nu}$ gives constraint that are zero only when $Q$ is constant or $B=0$. Unfortunately, this scenario is not possible in the case of neutron stars, as shown in figures~\ref{fig:ABSLy} and~\ref{fig:PQSLy}, which illustrate the profiles of $B(r)$ and $Q(r)$. Therefore, we calculated the constraint on $\mathring{\nabla}^\mu E^{\mu\nu}$ numerically. The results are shown in figure~\ref{fig:emu}. Here, we describe $f_{QQ}\Phi_r$, $f_{QQQ}\Psi_r$, and $\mathring{\nabla}_\mu E^{\mu\nu}$ for each model. The numerical plot illustrates that at the center and the surface of star, the constraint are zero due to the absence of nonmetricity. However, between the core and the surface, the constraint are non-zero but very small, ranging from the smallest order of $10^{-17}$ in the quadratic model to the largest order of $10^{-9}$ in the logarithmic model. From these results, it is still reasonable to assume $\mathring{\nabla}_\mu \mathcal{T}^{\mu\nu} \approx 0$, considering the constraint yields only very small values. Furthermore, in the neutron star calculations, particularly in the continuity equation, we focus on the matter part described by the EoS, which allows us to neglect these very small constraint term, thus enabling the use of the continuity equation in eq.~(\ref{eq12}) for neutron star calculations.

\section{Recovering TOV GR}\label{AppendixB}
The instability behaviour arises from our solutions, so we want to make sure that the TOV equations are capable of recovering to GR after setting $\alpha=0$. We will attempt to recover GR by two methods. In the first case, we will directly recover from eq. \eqref{eq11}, and in the second case, we will calculate the solution when $\alpha=0$ in eq. \eqref{eq15}. After setting $f(Q)=Q$ the eq. \eqref{eq11} become
\begin{eqnarray}\label{app1}
    \kappa e^A \rho&=& \frac{e^{A-B}}{2r^{2}}\left\{r^{2}e^{B}Q+\left[(e^{B}-1)(2+rA')+(1+e^{B})rB'\right]\right\},\nonumber\\
    \kappa  e^B p &=& \frac{-1}{2r^{2}}\left\{r^{2}e^{B}Q+\left[(e^{B}-1)(2+rA'+rB')-2rA'\right]\right\}.\nonumber\\
\end{eqnarray}
By subtitute the covariant $Q$, we will get
\begin{eqnarray}
   8\pi \rho r^{2} &=& 1 - e^{-B}(1-rB'),\label{app2}\\
    8\pi p r^{2} &=& -1 + e^{-B}(1+rA').\label{app3}
\end{eqnarray}
Integrating \eqref{app2}, we obtain:
\begin{equation}
e^{-B} = 1 - \frac{2m}{r}.
\end{equation}
Substituting this relation into \eqref{app2} and \eqref{app3}, and using the mass distribution in spherical coordinates, we get
\begin{equation}\label{app4}
A' = \frac{\frac{m}{r^2} + 4\pi p(r) r}{1 - \frac{2m}{r}}.
\end{equation}
Using the continuity equation from eq. \eqref{eq12}, we can derive the TOV equation in GR as:
\begin{equation}
\frac{dp}{dr} = \frac{(p + \rho)}{2m-r} \left( \frac{m}{r} + 4\pi p(r)r^2\right).
\end{equation}

\bibliographystyle{JCAP.bst}
\bibliography{Reference}

\providecommand{\href}[2]{#2}\begingroup\raggedright\begin{thebibliography}{10}

\bibitem{Riess1998}
A.G.~Riess, A.V.~Filippenko, P.~Challis, A.~Clocchiatti et~al.,
  \emph{{Observational Evidence from Supernovae for an Accelerating Universe
  and a Cosmological Constant}},
  \href{https://doi.org/10.1086/300499}{\emph{Astron. J.} {\bfseries 116}
  (1998) 1009}.

\bibitem{Perlmutter1999}
S.~Perlmutter, G.~Aldering, G.~Goldhaber, R.A.~Knop et~al., \emph{{Measurements
  of $\Omega$ and $\Lambda$ from 42 High-Redshift Supernovae}},
  \href{https://doi.org/10.1086/307221}{\emph{Astrophys. J.} {\bfseries 517}
  (1999) 565}.

\bibitem{Ade2016}
P.A.R.~Ade, N.~Aghanim, M.~Arnaud, M.~Ashdown et~al., \emph{{Planck 2015
  results: XIII. Cosmological parameters}},
  \href{https://doi.org/10.1051/0004-6361/201525830}{\emph{Astron. Astrophys.}
  {\bfseries 594} (2016) 63}.

\bibitem{Aghanim2016}
N.~Aghanim, M.~Arnaud, M.~Ashdown, J.~Aumont et~al., \emph{{Planck 2015
  results: XI. CMB power spectra, likelihoods, and robustness of parameters}},
  \href{https://doi.org/10.1051/0004-6361/201526926}{\emph{Astron. Astrophys.}
  {\bfseries 99} (2016) 594}.

\bibitem{Jimenez2019}
J.B.~Jim\'enez, L.~Heisenberg and T.S.~Koivisto, \emph{{The Geometrical Trinity
  of Gravity}}, \href{https://doi.org/10.3390/universe5070173}{\emph{Universe}
  {\bfseries 5} (2019) 173}.

\bibitem{Harada2020}
J.~Harada, \emph{{Connection independent formulation of general relativity}},
  \href{https://doi.org/10.1103/PhysRevD.101.024053}{\emph{Phys. Rev. D}
  {\bfseries 101} (2020) 024053}.

\bibitem{Aldrovandi2013}
R.~Aldrovandi and J.G.~Pereira, \emph{{Teleparallel Gravity}: {An
  Introduction}}, vol.~173, Springer (2013),
  \href{https://doi.org/10.1007/978-94-007-5143-9}{10.1007/978-94-007-5143-9}.

\bibitem{Maluf2013}
J.W.~Maluf, \emph{{The teleparallel equivalent of general relativity}},
  \href{https://doi.org/10.1002/andp.201200272}{\emph{Annalen Phys.} {\bfseries
  525} (2013) 339}.

\bibitem{Nester1998}
J.M.~Nester and H.-J.~Yo, \emph{{Symmetric teleparallel general relativity}},
  \href{https://doi.org/https://arxiv.org/abs/gr-qc/9809049}{\emph{Chinese J.
  Phys.} {\bfseries 37} (1999) 113}.

\bibitem{Adak2005}
M.~Adak, M.~Kalay and O.~Sert, \emph{{Lagrange formulation of the symmetric
  teleparallel gravity}},
  \href{https://doi.org/10.1142/S0218271806008474}{\emph{Int. J. Mod. Phys. D}
  {\bfseries 15} (2006) 619}.

\bibitem{Adak2008}
M.~Adak, O.~Sert, M.~Kalay and M.~Sari, \emph{{Symmetric Teleparallel Gravity:
  Some exact solutions and spinor couplings}},
  \href{https://doi.org/10.1142/S0217751X13501674}{\emph{Int. J. Mod. Phys. A}
  {\bfseries 28} (2013) 1350167}.

\bibitem{Mol2014}
I.~Mol, \emph{{The Non-Metricity Formulation of General Relativity}},
  \href{https://doi.org/10.1007/s00006-016-0749-8}{\emph{Adv. Appl. Clifford
  Algebras} {\bfseries 27} (2017) 2607}.

\bibitem{Jarv2018}
L.~J\"arv, M.~R\"unkla, M.~Saal and O.~Vilson, \emph{{Nonmetricity formulation
  of general relativity and its scalar-tensor extension}},
  \href{https://doi.org/10.1103/PhysRevD.97.124025}{\emph{Phys. Rev. D}
  {\bfseries 97} (2018) 124025}.

\bibitem{Jimenez2018}
J.B.~Jim\'enez, L.~Heisenberg and T.~Koivisto, \emph{{Coincident general
  relativity}}, \href{https://doi.org/10.1103/PhysRevD.98.044048}{\emph{Phys.
  Rev. D} {\bfseries 98} (2018) 044048}.

\bibitem{Jimenez2018a}
J.B.~Jim\'enez, L.~Heisenberg and T.S.~Koivisto, \emph{{Teleparallel Palatini
  theories}}, \href{https://doi.org/10.1088/1475-7516/2018/08/039}{\emph{JCAP}
  {\bfseries 08} (2018) 039}.

\bibitem{Gakis2020}
V.~Gakis, M.~Kr\ifmmode \check{s}\else \v{s}\fi{}\ifmmode~\check{s}\else
  \v{s}\fi{}\'ak, J.L.~Said and E.N.~Saridakis, \emph{{Conformal gravity and
  transformations in the symmetric teleparallel framework}},
  \href{https://doi.org/10.1103/PhysRevD.101.064024}{\emph{Phys. Rev. D}
  {\bfseries 101} (2020) 064024}.

\bibitem{Harko2018}
T.~Harko, T.S.~Koivisto, F.S.N.~Lobo, G.J.~Olmo and D.~Rubiera-Garcia,
  \emph{{Coupling matter in modified $Q$ gravity}},
  \href{https://doi.org/10.1103/PhysRevD.98.084043}{\emph{Phys. Rev. D}
  {\bfseries 98} (2018) 084043}.

\bibitem{Lazkoz2019}
R.~Lazkoz, F.S.N.~Lobo, M.~Ortiz-Ba\~nos and V.~Salzano, \emph{{Observational
  constraints of $f(Q)$ gravity}},
  \href{https://doi.org/10.1103/PhysRevD.100.104027}{\emph{Phys. Rev. D}
  {\bfseries 100} (2019) 104027}.

\bibitem{Lu2019}
J.~Lu, X.~Zhao and G.~Chee, \emph{{Cosmology in symmetric teleparallel gravity
  and its dynamical system}},
  \href{https://doi.org/10.1140/epjc/s10052-019-7038-3}{\emph{Eur. Phys. J. C}
  {\bfseries 79} (2019) 530}.

\bibitem{Jimenez2020}
J.B.~Jim\'enez, L.~Heisenberg, T.~Koivisto and S.~Pekar, \emph{{Cosmology in
  $f(Q)$ geometry}},
  \href{https://doi.org/10.1103/PhysRevD.101.103507}{\emph{Phys. Rev. D}
  {\bfseries 101} (2020) 103507}.

\bibitem{Barros2020}
B.J.~Barros, T.~Barreiro, T.~Koivisto and N.J.~Nunes, \emph{{Testing $F(Q)$
  gravity with redshift space distortions}},
  \href{https://doi.org/10.1016/j.dark.2020.100616}{\emph{Phys. Dark Universe}
  {\bfseries 30} (2020) 100616}.

\bibitem{Frusciante2021}
N.~Frusciante, \emph{{Signatures of $f(Q)$ gravity in cosmology}},
  \href{https://doi.org/10.1103/PhysRevD.103.044021}{\emph{Phys. Rev. D}
  {\bfseries 103} (2021) 044021}.

\bibitem{Anagnostopoulos2021}
F.K.~Anagnostopoulos, S.~Basilakos and E.N.~Saridakis, \emph{{First evidence
  that non-metricity $f(Q)$ gravity could challenge $\Lambda$CDM}},
  \href{https://doi.org/10.1016/j.physletb.2021.136634}{\emph{Phys. Lett. B}
  {\bfseries 822} (2021) 136634}.

\bibitem{Khyllep2021}
W.~Khyllep, A.~Paliathanasis and J.~Dutta, \emph{{Cosmological solutions and
  growth index of matter perturbations in $f(Q)$ gravity}},
  \href{https://doi.org/10.1103/PhysRevD.103.103521}{\emph{Phys. Rev. D}
  {\bfseries 103} (2021) 103521}.

\bibitem{Narawade2022}
S.A.~Narawade, L.~Pati, B.~Mishra and S.~Tripathy, \emph{{Dynamical system
  analysis for accelerating models in non-metricity $f(Q)$ gravity}},
  \href{https://doi.org/10.1016/j.dark.2022.101020}{\emph{Phys. Dark Universe}
  {\bfseries 36} (2022) 101020}.

\bibitem{Narawade2023a}
S.A.~Narawade and B.~Mishra, \emph{{Phantom Cosmological Model with
  Observational Constraints in $f(Q)$ Gravity}},
  \href{https://doi.org/10.1002/andp.202200626}{\emph{Ann. Phys.} {\bfseries
  535} (2023) 2200626}.

\bibitem{Narawade2023b}
S.A.~Narawade, S.P.~Singh and B.~Mishra, \emph{{Accelerating cosmological
  models in $f(Q)$ gravity and the phase space analysis}},
  \href{https://doi.org/10.1016/j.dark.2023.101282}{\emph{Phys. Dark Universe}
  {\bfseries 42} (2023) 101282}.

\bibitem{Heisenberg:2023lru}
L.~Heisenberg, \emph{{Review on f(Q) gravity}},
  \href{https://doi.org/10.1016/j.physrep.2024.02.001}{\emph{Phys. Rept.}
  {\bfseries 1066} (2024) 1}.

\bibitem{Heisenberg:2023wgk}
L.~Heisenberg, M.~Hohmann and S.~Kuhn, \emph{{Cosmological teleparallel
  perturbations}},
  \href{https://doi.org/10.1088/1475-7516/2024/03/063}{\emph{JCAP} {\bfseries
  03} (2024) 063}.

\bibitem{Nojiri:2024zab}
S.~Nojiri and S.D.~Odintsov, \emph{{Well-defined f(Q) gravity, reconstruction
  of FLRW spacetime and unification of inflation with dark energy epoch}},
  \href{https://doi.org/10.1016/j.dark.2024.101538}{\emph{Phys. Dark Universe}
  {\bfseries 45} (2024) 101538}.

\bibitem{Subramaniam2023}
G.~Subramaniam, A.~De, T.-H.~Loo and Y.K.~Goh, \emph{{How different connections
  in flat FLRW geometry impact energy conditions in $f(Q)$ theory?}},
  \href{https://doi.org/10.1002/prop.202300038}{\emph{Fortschritte der Phys.}
  {\bfseries 71} (2023) 2300038}.

\bibitem{Shabani2023}
H.~Shabani, A.~De and T.-H.~Loo, \emph{{Phase-space analysis of a novel
  cosmological model in $f(Q)$ theory}},
  \href{https://doi.org/10.1140/epjc/s10052-023-11722-5}{\emph{Eur. Phys. J. C}
  {\bfseries 83} (2023) 535}.

\bibitem{Paliathanasis2023}
A.~Paliathanasis, \emph{{Dynamical analysis of $f(Q)$-cosmology}},
  \href{https://doi.org/10.1016/j.dark.2023.101255}{\emph{Phys. Dark Universe}
  {\bfseries 41} (2023) 101255}.

\bibitem{Dimakis2022}
N.~Dimakis, A.~Paliathanasis, M.~Roumeliotis and T.~Christodoulakis,
  \emph{{FLRW solutions in $f(Q)$ theory: The effect of using different
  connections}}, \href{https://doi.org/10.1103/PhysRevD.106.043509}{\emph{Phys.
  Rev. D} {\bfseries 106} (2022) 043509}.

\bibitem{Heisenberg2023}
L.~Heisenberg, M.~Hohmann and S.~Kuhn, \emph{{Homogeneous and isotropic
  cosmology in general teleparallel gravity}},
  \href{https://doi.org/10.1140/epjc/s10052-023-11462-6}{\emph{Eur. Phys. J. C}
  {\bfseries 83} (2023) 315}.

\bibitem{Shabani2024}
H.~Shabani, A.~De, T.-H.~Loo and E.N.~Saridakis, \emph{{Cosmology of $f(Q)$
  gravity in non-flat Universe}},
  \href{https://doi.org/10.1140/epjc/s10052-024-12582-3}{\emph{Eur. Phys. J. C}
  {\bfseries 84} (2024) 285}.

\bibitem{Subramaniam2023a}
G.~Subramaniam, A.~De, T.-H.~Loo and Y.K.~Goh, \emph{{Energy condition bounds
  on $f(Q)$ model parameters in a curved FLRW Universe}},
  \href{https://doi.org/10.1016/j.dark.2023.101243}{\emph{Phys. Dark Universe}
  {\bfseries 41} (2023) 101243}.

\bibitem{Bhar2023}
P.~Bhar and J.M.Z.~Pretel, \emph{{Dark energy stars and quark stars within the
  context of $f(Q)$ gravity}},
  \href{https://doi.org/10.1016/j.dark.2023.101322}{\emph{Phys. Dark Universe}
  {\bfseries 42} (2023) 101322}.

\bibitem{Bhar2024}
P.~Bhar, K.N.~Singh, S.K.~Maurya and M.~Govender, \emph{{A four parameters
  quark star in quadratic $f(Q)$- action}},
  \href{https://doi.org/10.1016/j.dark.2023.101391}{\emph{Phys. Dark Universe}
  {\bfseries 43} (2024) 101391}.

\bibitem{Kaur2024}
S.~Kaur, S.K.~Maurya, S.~Shukla and B.~Dayanandan, \emph{{Charged anisotropic
  fluid sphere in $f(Q)$ gravity satisfying Vaidya-Tikekar metric}},
  \href{https://doi.org/10.1016/j.newast.2024.102230}{\emph{New Astron.}
  {\bfseries 110} (2024) 102230}.

\bibitem{Gul2024}
M.Z.~Gul, S.~Rani, M.~Adeel and A.~Jawad, \emph{{Viable and stable compact
  stars in $f({\mathcal{Q}})$ theory}},
  \href{https://doi.org/10.1140/epjc/s10052-023-12368-z}{\emph{Eur. Phys. J. C}
  {\bfseries 84} (2024) 8}.

\bibitem{Miller2019}
M.C.~Miller, F.K.~Lamb, A.J.~Dittmann, S.~Bogdanov et~al., \emph{{PSR
  J0030+0451 Mass and Radius from NICER Data and Implications for the
  Properties of Neutron Star Matter}},
  \href{https://doi.org/10.3847/2041-8213/ab50c5}{\emph{Astrophys. J. Lett.}
  {\bfseries 887} (2019) L24}.

\bibitem{Miller2021}
M.C.~Miller, F.K.~Lamb, A.J.~Dittmann, S.~Bogdanov et~al., \emph{{The Radius of
  PSR J0740+6620 from NICER and XMM-Newton Data}},
  \href{https://doi.org/10.3847/2041-8213/ac089b}{\emph{Astrophys. J. Lett.}
  {\bfseries 918} (2021) L28}.

\bibitem{Riley2021}
T.E.~Riley, A.L.~Watts, P.S.~Ray et~al., \emph{{A NICER View of the Massive
  Pulsar PSR J0740+6620 Informed by Radio Timing and XMM-Newton Spectroscopy}},
  \href{https://doi.org/10.3847/2041-8213/ac0a81}{\emph{Astrophys. J. Lett.}
  {\bfseries 918} (2021) L27}.

\bibitem{Abbott2017}
B.P.~Abbott, R.~Abbott, T.D.~Abbott, F.~Acernese et~al., \emph{{GW170817:
  Observation of Gravitational Waves from a Binary Neutron Star Inspiral}},
  \href{https://doi.org/10.1103/physrevlett.119.161101}{\emph{Phys. Rev. Lett.}
  {\bfseries 119} (2017) 161101}.

\bibitem{Abbott2020}
R.~Abbott, T.D.~Abbott, S.~Abraham, F.~Acernese et~al., \emph{{GW190814:
  Gravitational Waves from the Coalescence of a 23 Solar Mass Black Hole with a
  2.6 Solar Mass Compact Object}},
  \href{https://doi.org/10.3847/2041-8213/ab960f}{\emph{Astrophys. J. Lett.}
  {\bfseries 896} (2020) L44}.

\bibitem{Lattimer2016}
J.M.~Lattimer and M.~Prakash, \emph{{The equation of state of hot, dense matter
  and neutron stars}},
  \href{https://doi.org/10.1016/j.physrep.2015.12.005}{\emph{Phys. Rep.}
  {\bfseries 621} (2016) 127}.

\bibitem{Hebeler2013}
K.~Hebeler, J.M.~Lattimer, C.J.~Pethick and A.~Schwenk, \emph{{Equation of
  State and Neutron Star Properties Constrained by Nuclear Physics and
  Observation}},
  \href{https://doi.org/10.1088/0004-637X/773/1/11}{\emph{Astrophys. J.}
  {\bfseries 773} (2013) 11}.

\bibitem{Ozel2016}
F.~\"Ozel and P.~Freire, \emph{{Masses, Radii, and the Equation of State of
  Neutron Stars}},
  \href{https://doi.org/10.1146/annurev-astro-081915-023322}{\emph{Annu. Rev.
  Astron. Astrophys.} {\bfseries 54} (2016) 401}.

\bibitem{Steiner2017}
A.W.~Steiner, C.O.~Heinke, S.~Bogdanov, C.K.~Li, W.C.G.~Ho, A.~Bahramian
  et~al., \emph{{Constraining the mass and radius of neutron stars in globular
  clusters}}, \href{https://doi.org/10.1093/mnras/sty215}{\emph{Mon. Notices
  Royal Astron. Soc.} {\bfseries 476} (2018) 421}.

\bibitem{Bertotti1969}
B.~Bertotti, A.~Cavaliere and F.~Pacini, \emph{{Rotating Neutron Stars and
  Pulsar Emission}}, \href{https://doi.org/10.1038/221624A0}{\emph{nature}
  {\bfseries 221} (1969) 624}.

\bibitem{Steiner2013}
A.W.~Steiner, J.M.~Lattimer and E.F.~Brown, \emph{{THE NEUTRON STAR MASS-RADIUS
  RELATION AND THE EQUATION OF STATE OF DENSE MATTER}},
  \href{https://doi.org/10.1088/2041-8205/765/1/l5}{\emph{Astrophys. J.}
  {\bfseries 765} (2013) L5}.

\bibitem{Steiner2014}
A.W.~Steiner, S.~Gandolfi, F.J.~Fattoyev and W.G.~Newton, \emph{{Using neutron
  star observations to determine crust thicknesses, moments of inertia, and
  tidal deformabilities}},
  \href{https://doi.org/10.1103/PhysRevC.91.015804}{\emph{Phys. Rev. C}
  {\bfseries 91} (2015) 015804}.

\bibitem{Chandrasekhar1931}
S.~Chandrasekhar, \emph{{The Maximum Mass of Ideal White Dwarfs}},
  \href{https://doi.org/10.1086/143324}{\emph{Astrophys. J.} {\bfseries 74}
  (1931) 81}.

\bibitem{Rawls2011}
M.L.~Rawls, J.A.~Orosz, J.E.~McClintock, M.A.P.~Torres, C.D.~Bailyn and
  M.M.~Buxton, \emph{{REFINED NEUTRON STAR MASS DETERMINATIONS FOR SIX
  ECLIPSING X-RAY PULSAR BINARIES*}},
  \href{https://doi.org/10.1088/0004-637X/730/1/25}{\emph{Astrophys. J.}
  {\bfseries 730} (2011) 25}.

\bibitem{Mullally2009}
F.~Mullally, C.~Badenes, S.E.~Thompson and R.~Lupton, \emph{{TWINS: THE TWO
  SHORTEST PERIOD NON-INTERACTING DOUBLE DEGENERATE WHITE DWARF STARS}},
  \href{https://doi.org/10.1088/0004-637X/707/1/L51}{\emph{Astrophys. J.}
  {\bfseries 707} (2009) L51}.

\bibitem{Demorest2010}
P.B.~Demorest, T.~Pennucci, S.M.~Ransom, M.S.E.~Roberts and J.W.T.~Hessels,
  \emph{{A two-solar-mass neutron star measured using Shapiro delay}},
  \href{https://doi.org/10.1038/nature09466}{\emph{nature} {\bfseries 467}
  (2010) 1081}.

\bibitem{Zhang2019}
N.-B.~Zhang and B.-A.~Li, \emph{{Implications of the Mass
  $M={2.17}_{-0.10}^{+0.11}$ $M_\odot$ of PSR J0740+6620 on the Equation of
  State of Super-dense Neutron-rich Nuclear Matter}},
  \href{https://doi.org/10.3847/1538-4357/ab24cb}{\emph{Astrophys. J.}
  {\bfseries 879} (2019) 99}.

\bibitem{Ganguly2014}
A.~Ganguly, R.~Gannouji, R.~Goswami and S.~Ray, \emph{{Neutron stars in the
  Starobinsky model}},
  \href{https://doi.org/10.1103/physrevd.89.064019}{\emph{Phys. Rev. D}
  {\bfseries 89} (2014) 064019}.

\bibitem{Astashenok:2014nua}
A.V.~Astashenok, S.~Capozziello and S.D.~Odintsov, \emph{{Extreme neutron stars
  from Extended Theories of Gravity}},
  \href{https://doi.org/10.1088/1475-7516/2015/01/001}{\emph{JCAP} {\bfseries
  01} (2015) 001}.

\bibitem{Yazadjiev2014}
S.S.~Yazadjiev, D.D.~Doneva, K.D.~Kokkotas and K.V.~Staykov,
  \emph{{Non-perturbative and self- consistent models of neutron stars in
  $R$-squared gravity}},
  \href{https://doi.org/10.1088/1475-7516/2014/06/003}{\emph{JCAP} {\bfseries
  06} (2014) 003}.

\bibitem{Capozziello:2015yza}
S.~Capozziello, M.~De~Laurentis, R.~Farinelli and S.D.~Odintsov,
  \emph{{Mass-radius relation for neutron stars in f(R) gravity}},
  \href{https://doi.org/10.1103/PhysRevD.93.023501}{\emph{Phys. Rev. D}
  {\bfseries 93} (2016) 023501}.

\bibitem{Astashenok:2017dpo}
A.V.~Astashenok, S.D.~Odintsov and A.~de~la Cruz-Dombriz, \emph{{The realistic
  models of relativistic stars in $f(R) = R + \alpha R^2$ gravity}},
  \href{https://doi.org/10.1088/1361-6382/aa8971}{\emph{Class. Quant. Grav.}
  {\bfseries 34} (2017) 205008}.

\bibitem{Kpadonou2016}
A.V.~Kpadonou, M.J.S.~Houndjo and M.E.~Rodrigues,
  \emph{{Tolman-Oppenheimer-Volkoff equations and their implications for the
  structures of relativistic stars in $f(T)$ gravity}},
  \href{https://doi.org/10.1007/s10509-016-2805-1}{\emph{Astrophys. and Space
  Sci.} {\bfseries 361} (2016) 244}.

\bibitem{Pace2017}
M.~Pace and J.L.~Said, \emph{{A perturbative approach to neutron stars in $f(T,
  \mathcal {T})$-gravity}},
  \href{https://doi.org/10.1140/epjc/s10052-017-4838-1}{\emph{Eur. Phys. J. C}
  {\bfseries 77} (2017) 283}.

\bibitem{Fortes2022}
H.G.M.~Fortes and J.C.N.~Araujo, \emph{{Solving Tolman-Oppenheimer-Volkoff
  equations in $f(T)$ gravity: a novel approach}},
  \href{https://doi.org/10.1088/1361-6382/aca384}{\emph{Class. Quant. Grav.}
  {\bfseries 39} (2022) 245017}.

\bibitem{Araujo2022}
J.C.N.~de~Araujo and H.G.M.~Fortes, \emph{{Solving
  Tolman\textendash{}Oppenheimer\textendash{}Volkoff equations in $f(T)$
  gravity: A novel approach applied to some realistic equations of state}},
  \href{https://doi.org/10.1142/S0218271822501012}{\emph{Int. J. Mod. Phys. D}
  {\bfseries 31} (2022) 2250101}.

\bibitem{Lin2021}
R.-H.~Lin and X.-H.~Zhai, \emph{{Spherically symmetric configuration in $f(Q)$
  gravity}}, \href{https://doi.org/10.1103/PhysRevD.103.124001}{\emph{Phys.
  Rev. D} {\bfseries 103} (2021) 124001}.

\bibitem{Zhao2022}
D.~Zhao, \emph{{Covariant formulation of $f(Q)$ theory}},
  \href{https://doi.org/10.1140/epjc/s10052-022-10266-4}{\emph{Eur. Phys. J. C}
  {\bfseries 82} (2022) 303}.

\bibitem{Beh2022}
J.-T.~Beh, T.-H.~Loo and A.~De, \emph{{Geodesic deviation equation in
  $f(Q)$-gravity}},
  \href{https://doi.org/10.1016/j.cjph.2021.11.026}{\emph{Chinese J. Phys.}
  {\bfseries 77} (2022) 1551}.

\bibitem{De2023}
A.~De and T.-H.~Loo, \emph{{On the viability of $f(Q)$ gravity models}},
  \href{https://doi.org/10.1088/1361-6382/accef7}{\emph{Classical and Quantum
  Gravity} {\bfseries 40} (2023) 115007}.

\bibitem{Sokoliuk2023}
O.~Sokoliuk, S.~Arora, S.~Praharaj, A.~Baransky and P.K.~Sahoo, \emph{{On the
  impact of $f(Q)$ gravity on the large scale structure}},
  \href{https://doi.org/10.1093/mnras/stad968}{\emph{Mon. Notices Royal Astron.
  Soc.} {\bfseries 522} (2023) 252}.

\bibitem{Najera2023}
J.A.~N\'ajera, C.A.~Alvarado and C.~Escamilla-Rivera, \emph{{Constraints on
  $f(Q)$ logarithmic model using gravitational wave standard sirens}},
  \href{https://doi.org/10.1093/mnras/stad2180}{\emph{Mon. Notices Royal
  Astron. Soc.} {\bfseries 524} (2023) 5280}.

\bibitem{Israel1966}
W.~Israel, \emph{{Singular hypersurfaces and thin shells in general
  relativity}}, \href{https://doi.org/10.1007/BF02710419}{\emph{Nuovo cimento
  B} {\bfseries 44S10} (1966) 1}.

\bibitem{Marolf2005}
D.~Marolf and S.~Yaida, \emph{{Energy conditions and junction conditions}},
  \href{https://doi.org/10.1103/physrevd.72.044016}{\emph{Phys. Rev. D}
  {\bfseries 72} (2005) 044016}.

\bibitem{Deruelle2008}
N.~Deruelle, M.~Sasaki and Y.~Sendouda, \emph{{Junction Conditions in $f(R)$
  Theories of Gravity}}, \href{https://doi.org/10.1143/ptp.119.237}{\emph{Prog.
  theor. phys.} {\bfseries 119} (2008) 237}.

\bibitem{Senovilla2013}
J.M.M.~Senovilla, \emph{{Junction conditions for $F(R)$-gravity and their
  consequences}}, \href{https://doi.org/10.1103/physrevd.88.064015}{\emph{Phys.
  Rev. D} {\bfseries 88} (2013) 064015}.

\bibitem{Feng2017}
W.-X.~Feng, C.-Q.~Geng, W.F.~Kao and L.-W.~Luo, \emph{{Equation-of-state of
  neutron stars with junction conditions in the Starobinsky model}},
  \href{https://doi.org/10.1142/s0218271817501863}{\emph{Int. J. of Modd. Phys.
  D} {\bfseries 27} (2017) 1750186}.

\bibitem{Maurya2022}
S.K.~Maurya, K.N.~Singh, S.V.~Lohakare and B.~Mishra, \emph{{Anisotropic
  Strange Star Model Beyond Standard Maximum Mass Limit by Gravitational
  Decoupling in $f(Q)$ Gravity}},
  \href{https://doi.org/10.1002/prop.202200061}{\emph{Fortschritte der Phys.}
  {\bfseries 70} (2022) 2200061}.

\bibitem{Maurya2023}
S.K.~Maurya, K.N.~Singh et~al., \emph{{The Effect of Gravitational Decoupling
  on Constraining the Mass and Radius for the Secondary Component of GW190814
  and Other Self-bound Strange Stars in $f(Q)$ Gravity Theory}},
  \href{https://doi.org/10.3847/1538-4365/ad0154}{\emph{Astrophys. J., Suppl.
  Ser.} {\bfseries 269} (2023) 35}.

\bibitem{Chaudharya2024}
S.~Chaudharya, S.K.~Maurya, J.~Kumara and G.~Mustafa, \emph{{Most general
  isotropic charged fluid solution for Buchdahl model in $\mathscr{F}(Q)$
  gravity}},  \href{https://arxiv.org/abs/arXiv:2406.18604}{{\ttfamily
  arXiv:2406.18604}}.

\bibitem{Douchin2001}
F.~Douchin and P.~Haensel, \emph{{A unified equation of state of dense matter
  and neutron star structure}},
  \href{https://doi.org/10.1051/0004-6361:20011402}{\emph{Astron. Astrophys.}
  {\bfseries 380} (2001) 151}.

\bibitem{Potekhin2013}
A.Y.~Potekhin, A.F.~Fantina, N.~Chamel, J.M.~Pearson and S.~Goriely,
  \emph{{Analytical representations of unified equations of state for
  neutron-star matter}},
  \href{https://doi.org/10.1051/0004-6361/201321697}{\emph{Astron. Astrophys.}
  {\bfseries 560} (2013) A48}.

\bibitem{APR4}
A.~Akmal, V.R.~Pandharipande and D.G.~Ravenhall, \emph{{Equation of state of
  nucleon matter and neutron star structure}},
  \href{https://doi.org/10.1103/PhysRevC.58.1804}{\emph{Phys. Rev. C}
  {\bfseries 58} (1998) 1804}.

\bibitem{MS1b}
H.~M\"uller and B.D.~Serot, \emph{{Relativistic mean-field theory and the
  high-density nuclear equation of state}},
  \href{https://doi.org/10.1016/0375-9474(96)00187-x}{\emph{Nucl. Phys. A}
  {\bfseries 606} (1996) 508}.

\bibitem{Read2009}
J.S.~Read, B.D.~Lackey, B.J.~Owen and J.L.~Friedman, \emph{{Constraints on a
  phenomenologically parametrized neutron-star equation of state}},
  \href{https://doi.org/10.1103/PhysRevD.79.124032}{\emph{Phys. Rev. D}
  {\bfseries 79} (2009) 124032}.

\bibitem{Linares2018}
M.~Linares, T.~Shahbaz and J.~Casares, \emph{{Peering into the Dark Side:
  Magnesium Lines Establish a Massive Neutron Star in PSR J2215+5135}},
  \href{https://doi.org/10.3847/1538-4357/aabde6}{\emph{Astrophys. J.}
  {\bfseries 859} (2018) 54}.

\bibitem{Riley2019}
T.E.~Riley, A.L.~Watts, S.~Bogdanov et~al., \emph{{A NICER View of PSR
  J0030+0451: Millisecond Pulsar Parameter Estimation}},
  \href{https://doi.org/10.3847/2041-8213/ab481c}{\emph{Astrophys. J. Lett.}
  {\bfseries 887} (2019) L21}.

\bibitem{Ganiou2017}
M.G.~Ganiou, C.~A\"\i{}namon, M.J.S.~Houndjo and J.~Tossa, \emph{{Strong
  magnetic field effects on neutron stars within $f(T)$ theory of gravity}},
  \href{https://doi.org/10.1140/epjp/i2017-11499-3}{\emph{Eur. Phys. J. Plus}
  {\bfseries 132} (2017) 250}.

\bibitem{Iliji2018}
S.c.v.~Iliji\ifmmode~\acute{c}\else \'{c}\fi{} and M.~Sossich, \emph{{Compact
  stars in extended theory of gravity}},
  \href{https://doi.org/10.1103/physrevd.98.064047}{\emph{Phys. Rev. D}
  {\bfseries 98} (2018) }.

\bibitem{Buch1959}
H.A.~Buchdahl, \emph{General relativistic fluid spheres},
  \href{https://doi.org/10.1103/PhysRev.116.1027}{\emph{Phys. Rev.} {\bfseries
  116} (1959) 1027}.

\bibitem{N.Straumann}
V.~M\"uller, \emph{{N. Straumann: General relativity and relativistic
  astrophysics. Springer-Verlag, Berlin, Heidelberg, New York, Tokyo 1984. XIII
  + 459 Seiten. DM 112,-}},
  \href{https://doi.org/https://doi.org/10.1002/asna.2113080106}{\emph{Astronomische
  Nachrichten} {\bfseries 308} (1987) 40}.

\bibitem{Barker2024}
W.~Barker and S.~Zell, \emph{{Consistent particle physics in metric-affine
  gravity from extended projective symmetry}},
  \href{https://arxiv.org/abs/arXiv:2402.14917}{{\ttfamily arXiv:2402.14917}}.

\bibitem{Lohakare2023}
S.V.~Lohakare, S.K.~Maurya, K.N.~Singh, B.~Mishra and A.~Errehymy,
  \emph{{Influence of three parameters on maximum mass and stability of strange
  star under linear $f(Q)$ \ensuremath{-} action}},
  \href{https://doi.org/10.1093/mnras/stad2861}{\emph{Mon. Notices Royal
  Astron. Soc.} {\bfseries 526} (2023) 3796}.

\bibitem{Lin2021a}
R.-H.~Lin, X.-N.~Chen and X.-H.~Zhai, \emph{{Realistic neutron star models in
  $f(T)$ gravity}},
  \href{https://doi.org/10.1140/epjc/s10052-022-10268-2}{\emph{Eur. Phys. J. C}
  {\bfseries 82} (2022) 308}.

\bibitem{Astashenok2019}
A.V.~Astashenok, A.S.~Baigashov and S.A.~Lapin, \emph{{Neutron stars in frames
  of $R^{2}-$gravity and gravitational waves}},
  \href{https://doi.org/10.1142/s021988781950004x}{\emph{Int. J. Geom. Methods
  Mod. Phys.} {\bfseries 16} (2019) 1950004}.

\bibitem{Feola2020}
P.~Feola, X.J.~Forteza, S.~Capozziello, R.~Cianci and S.~Vignolo,
  \emph{{Mass-radius relation for neutron stars in
  $f(R)=R+\ensuremath{\alpha}{R}^{2}$ gravity: A comparison between purely
  metric and torsion formulations}},
  \href{https://doi.org/10.1103/PhysRevD.101.044037}{\emph{Phys. Rev. D}
  {\bfseries 101} (2020) 044037}.

\bibitem{Numajiri2023}
K.~Numajiri, Y.-X.~Cui, T.~Katsuragawa and S.~Nojiri, \emph{{Revisiting compact
  star in $F(R)$ gravity: Roles of chameleon potential and energy conditions}},
  \href{https://doi.org/10.1103/PhysRevD.107.104019}{\emph{Phys. Rev. D}
  {\bfseries 107} (2023) 104019}.

\bibitem{D'Ambrosio2022}
F.~D'Ambrosio, S.D.B.~Fell, L.~Heisenberg and S.~Kuhn, \emph{{Black holes in
  $f(Q)$ gravity}},
  \href{https://doi.org/10.1103/PhysRevD.105.024042}{\emph{Phys. Rev. D}
  {\bfseries 105} (2022) 024042}.

\end{thebibliography}\endgroup

\end{document}